\documentclass[12pt]{article}
\usepackage{epsfig,psfrag,cite,float}
\def \gh{\vphantom{$\fbox{\Big[}$}}

\def \beq{\begin{equation}}
\def \eeq{\end{equation}}
\def \bea{\begin{eqnarray}}
\def \eea{\end{eqnarray}}
\def \ben{\begin{enumerate}}
\def \een{\end{enumerate}}
\def \bit{\begin{itemize}}
\def \eit{\end{itemize}}

\def \branch{{\cal B}}
\def \eff{\hbox{eff}}
\def \Im{{\hbox{Im}}\,}
\def \Re{{\hbox{Re}}\,}
\def \gev{{\hbox{GeV}}}

\def \tev{{\hbox{TeV}}}

\def \cl#1{{#1\%\ \mathrm{C.L.}}}

\def \dis{\displaystyle}

\def \eq#1{Eq.~(\ref{#1})}

\def \fig#1{Fig.~\ref{#1}}

\def \nn{\nonumber}
\def \rf{Ref.~\cite}

\def \a{\alpha}
\def \b{\beta}

\def \g{\gamma}

\def \d{\delta}

\def \l{\lambda}

\def \m{\mu}
\def \n{\nu}
\def \o{\omega}

\def \p{\pi}

\def \t{\tau}

\def \alphas{\alpha_s}

\begin{document}
\begin{titlepage}

\begin{flushright}
DESY 01-217\\
BUTP-01-21\\
SLAC-PUB-9076\\
\end{flushright} 

\begin{center} 
\bf \large
Improved Model-Independent Analysis of Semileptonic \\ 
and Radiative Rare $B$ Decays
\end{center}

\bigskip

\begin{center}
\large
A. Ali\footnote{E-mail address: ali@mail.desy.de} ~and
E. Lunghi\footnote{E-mail address: lunghi@mail.desy.de}
\end{center}

\begin{center}
Deutsches Elektronen Synchrotron, DESY, \\
Notkestrasse 85, D-22607 Hamburg, Germany
\end{center}

\begin{center}
\large
C. Greub\footnote{E-mail address: greub@itp.unibe.ch}\footnote{Work
partially supported by Schweizerischer Nationalfonds} 
\end{center}

\begin{center}
Institut f\"ur Theoretische Physik, Universit\"at Bern \\
CH-3012 Bern, Switzerland
\end{center}

\begin{center}
\large
G. Hiller\footnote{E-mail address: ghiller@SLAC.stanford.EDU}\footnote{Work
supported by the Department of Energy, Contract
DE-AC03-76SF00515.}
\end{center}

\begin{center}
Stanford Linear Accelerator Center, Stanford University, Stanford,\\
CA 94309, USA
\end{center}

\bigskip
\begin{abstract}
We update the branching ratios for the inclusive decays $B\to X_s
\ell^+ \ell^-$ and the exclusive decays $B\to (K,K^*) \ell^+ \ell^-$,
with $\ell=e,\; \m$, in the standard model by including the explicit
$O(\a_s)$ and $\Lambda_{\hbox{\tiny QCD}} / m_b$ corrections. This
framework is used in conjunction with the current measurements
of the branching ratios for $B\to X_s \g$ and $B\to K \ell^+ \ell^-$
decays and upper limits on the branching ratios for the decays
$B\to (K^*,X_s) \ell^+ \ell^-$ to work out bounds on the Wilson
coefficients $C_7$, $C_8$, $C_9$ and $C_{10}$ appearing in the effective
Hamiltonian formalism. The resulting bounds are found to be
consistent with the predictions of the standard model and some
variants of supersymmetric theories. We illustrate the constraints on
supersymmetric parameters that the current data on rare $B$ decays implies
in the context of minimal flavour violating model and in more general
scenarios admitting additional flavour changing mechanisms. Precise
measurements of the dilepton invariant mass distributions in the decays
$B \to (X_s,K^*,K) \ell^+ \ell^-$, in particular in the lower dilepton
mass region, and the forward-backward
asymmetry in the decays $B \to (X_s,K^*) \ell^+ \ell^-$, will greatly help
in discriminating among the SM and various supersymmetric theories.   
\end{abstract}
\end{titlepage}

\section{Introduction}
\label{intro}
The measurement of the inclusive decay $B \to X_s \gamma$, first
reported by the CLEO collaboration in 1995 \cite{Alam:1995aw}, has
received a resounding reception in the interested theoretical physics
community, both as a precision test of the standard model in the
flavour sector and as a harbinger of new physics, in particular
supersymmetry \cite{Greub:1999sv}. In the meanwhile, the 
branching ratio for this decay has become quite precise through the
subsequent measurements by the CLEO \cite{cleobsg}, ALEPH \cite{alephbsg}
and BELLE \cite{bellebsg} collaborations, with the BABAR measurements
keenly awaited. The present experimental average of the branching
ratio ${\cal B}(B \to X_s \gamma)= (3.22 \pm 0.40)\times 10^{-4}$ is
in good agreement with the next-to-leading order prediction 
of the same in the standard model (SM), estimated as ${\cal B}(B \to X_s
\gamma)_{\rm SM}= (3.35 \pm 0.30)\times 10^{-4}$
\cite{Chetyrkin:1997vx,Kagan:1999ym} for the pole quark mass ratio
$m_c/m_b=0.29 \pm 0.02$, rising to ${\cal B}(B \to X_s \gamma)_{\rm
SM}= (3.73 \pm 0.30)\times 10^{-4}$ \cite{Gambino:2001ew}, if one uses
the input value $m_c^{\overline{\rm MS}}(\mu)/m_{b, pole}=0.22 \pm
0.04$, where $m_c^{\overline{\rm MS}}(\mu)$ is the charm quark mass in
the $\overline{\rm MS}$-scheme, evaluated at a scale $\mu$ in the range
$m_c < \mu < m_b$. The inherent uncertainty reflects in part the present
accuracy of the theoretical branching ratio, which is limited
to $O({\alpha_s})$, and in part the imprecise measurements of the
photon energy spectrum in $B \to X_s \gamma$ decays. Despite the current
theoretical dispersion on the branching ratio, the agreement between
experiment and the SM is quite impressive and this has been used to put
non-trivial constraints on the parameters of models incorporating
beyond-the-SM physics, in particular supersymmetry (see, for example, 
Ref.~\cite{Everett:2001yy} for a recent analysis in a 
supersymmetric scenario).
While the measurement of the decay $B \to X_s \gamma$ is being
consolidated, several other radiative
and semileptonic rare $B$-decays are being searched for. In
particular, first measurements of semileptonic rare $B$-decays have
been recently reported in the $B \to K \mu^+\mu^-$ and $B \to K e^+
e^-$ modes by the BELLE collaboration \cite{bellebsll}, and upper
limits have been put in a number of other related decay modes
\cite{bellebsll,babarbsll,Affolder:1999eb,Anderson:2001nt}.  The
current measurements of the exclusive modes are in agreement with the
expectations in the SM \cite{Ali:1997bm,Ali:2000mm}, calculated in
next-to-leading logarithmic (NLO) approximation, taking into account the
experimental and theoretical errors. This, for example, can be judged from
the comparison of the combined
branching ratio for the decay modes $B \to K \ell^+ \ell^-$,
$\ell=e,\mu$, reported by the BELLE collaboration ${\cal B}(B \to K
\ell^+ \ell^-)=0.75 ^{+0.25}_{-0.21} \pm 0.09 \times 10^{-6}$ with the
Light-cone QCD sum rule based estimates of the same,
${\cal B}(B \to K \ell^+ \ell^-)= 0.57 ^{+0.16}_{-0.10} \times
10^{-6}$ \cite{Ali:2000mm}. The upper limits on the inclusive decays
$B \to X_s \ell^+ \ell^-$ and the exclusive decays $B \to K^* \ell^+
\ell^-$ are now approaching their respective
SM-based estimates, as we also show quantitatively in this paper.

With increased statistical power of experiments at the $B$
factories in the next several years, the decays discussed above  and
related rare $B$ decays
will be measured very precisely. On the theoretical side, partial
results in next-to-next-to-leading logarithmic (NNLO) accuracy are now
available in the inclusive decays $B \to X_s \ell^+ \ell^-$
\cite{BMU,AAGW}. Recalling that the lowest order contribution for these
decays starts at $O(1/\alpha_s)$, as opposed to the decay $B \to X_s \g$,
which starts at $O(\alpha_s^0)$, the NNLO accuracy in  $B \to X_s \ell^+
\ell^-$ amounts to calculating explicit $O(\alpha_s)$ improvements. The
same accuracy in $\alpha_s$ amounts to calculating the decay
$B \to X_s \g$ in NLO. We also recall that power corrections in $\Lambda_{\rm
QCD}/m_b$ \cite{Ali:1998sf} and in $\Lambda_{\rm QCD}/m_c$
\cite{Buchalla:1997ky} are also known. What concerns the exclusive
decays, some theoretical progress in calculating their decay rates
to NLO accuracy in the $B \to (K^*,\rho) \gamma$
\cite{Ali:2001ez,Beneke:2001at,Bosch:2001gv}, and to NNLO accuracy in $B
\to K^* \ell^+ \ell^-$ \cite{Beneke:2001at} decays, including the leading
$\Lambda_{\rm QCD}/M_B$, has been reported.  Comparisons of
these theoretical estimates with data on $B \to K^* \gamma$ decays
\cite{cleobsg,TajimaH:2001,Aubert:2001} have led to important
inferences on the magnetic moment form factor.  Our purpose in this
paper is to incorporate these theoretical improvements, carried out in
the context of the SM, and phenomenological implications from the 
observed radiative decays, and examine  the
quantitative {\it rapport} between the SM and current measurements of the
semileptonic rare $B$-decays.  An equally important undertaking of our
analysis is to investigate the impact of the current
experimental measurements on the parameters of the possible
supersymmetric extensions of the SM. The question which we address in
this context is the following: do the current measurements in
semileptonic rare $B$-decays already provide more restrictive constraints
on the parameters of the supersymmetric models than are provided by the
$B \to X_s \gamma$ measurements? We find that the decays $B \to
(X_s,K^*,K) \ell^+ \ell^-$ do provide additional constraints in some
parts of the supersymmetric space, though with the current experimental
knowledge the decay $B \to X_s \gamma$ remains more restrictive over most
of the supersymmetric space. This is expected to change with improved
precision on the semileptonic rare $B$-decays, which we illustrate in a
number of supersymmetric scenarios. 

Our analysis is carried out in the effective Hamiltonian approach,
obtained by integrating out the heavy
degrees of freedom, defined below (see Eq.~(\ref{Heff})). However, we
make the tacit assumption that the dominant effects of an underlying
supersymmetric theory can be implemented by using the SM operator
basis for the effective Hamiltonian. Thus, supersymmetric effects enter
in our analysis through the modifications of the Wilson coefficients
which in the SM are calculated at some high scale, denoted generically
by $\m_W$, with the SM anomalous dimension matrix controlling the
renormalization of these coefficients to a lower scale, typically
$\mu_b=O(m_b)$. Restricting the operator basis to the one in the SM
obviously does not cover the most general supersymmetric case, but we
think that it covers an important part of the underlying parameter space,
and hence can be employed to undertake searches for supersymmetric effects
in rare $B$-decays. Within this operator basis,
we have split our analysis in two parts. In the first part we update
the branching ratios for the decays $B\to (X_s,K,K^*) \ell^+\ell^-$,
$\ell = e, \; \m$, in the standard model. In doing this, we work out
the parametric uncertainties due to the scale--dependence, top quark mass,
$m_t$, and the ratio of the quark masses $m_c/m_b$. Combining the
individual errors $\delta{\cal B}(\m)$,
$\delta{\cal B}(m_t)$, and $\delta{\cal B}(m_c/m_b)$ in quadrature,
we find that the resulting theoretical uncertainties are $\d \branch
(B\to X_s e^+e^-) \simeq \pm 15 \%$ and $\d \branch (B\to X_s
\m^+\m^-) \simeq \pm 17 \%$. The corresponding theoretical
uncertainties on the exclusive decay branching ratios are larger,
due to the form factors, and estimated at typically $O(\pm 35\%)$. Using
this updated theoretical framework,
we extract model-independent constraints that current data (summarized
below in Eqs.~(\ref{bsgexp})--(\ref{bseeexp})) provides on the Wilson
coefficients $C_{7}$ -- $C_{10}$, which appear in the effective
Hamiltonian. We first work out
the constraints on $C_7$ and $C_8$ implied by the $B\to X_s \g$
measurement. To that end, we define the quantities
$R_{7,8}(\m_W)\equiv C_{7,8}^{\rm tot}(\m_W)/C_{7,8}(\m_W)$, and work out
bounds on 
them. Data on $B \to X_s \g$ allows both $R_{7,8}(\m_W)>0$
and $R_{7,8}(\m_W)<0$ solutions, which we show in terms of the allowed
regions in the $(R_7(\m_W)$,$R_8(\m_W))$ and $(R_7(\mu_b)$,$R_8(\mu_b))$
planes.
We then transcribe the impact of the $B\to (X_s,K,K^*)
\ell^+ \ell^-$ experimental data on the allowed regions in
the $[C_9,C_{10}]$ plane. Depending on the two branches for the quantities
$R_{7,8}(\m_W)$, we display the constraints in terms of $C_9^{\rm
NP}(\m_W)$ and $C_{10}^{\rm NP}$. We show that the SM solution
(corresponding to the point $(0,0)$ in this plane for the case 
$R_{7,8}(\m_W)=1$) is allowed by present data. More importantly, from
the point of view of supersymmetry, our analysis shows the allowed
region in the $[C_9,C_{10}]$ plane, which leaves considerable room for
beyond-the-SM contributions to these quantities. In fact, in some
allowed regions, phenomenological profiles of semileptonic rare $B$-decays
can measurably differ from the corresponding ones in the SM.

The second part of our supersymmetric analysis deals with specific SUSY
models and we quantify the additional constraints that the generic $b\to s
\ell^+ \ell^-$ data implies for the parameters of these models. We
show that no useful bounds beyond what are already known from the $B
\to X_s \gamma$ analysis are at present obtained in the so--called
minimal flavour violating models (including the constrained minimal
supersymmetric standard model MSSM) \cite{Ciuchini:1998xy}. This
reflects the generically small deviations to the SM rates and
distributions anticipated in these models, as the allowed
supersymmetric parameters are already highly constrained.  Working in
the mass insertion approximation \cite{mia1}, we show that
insertions in the down--squark sector (that enter principally through the
gluino penguin and box diagrams) are not constrained either from present
data. On the other
hand, insertions in the up--squark sector get in some parts of the
SUSY parameter space genuinely new constraints. To show possible
supersymmetric effects that precise measurements in semileptonic rare
$B$-decays may reveal, we work out the forward-backward asymmetry in
$B \to X_s \ell^+ \ell^-$ for four illustrative points in the
$(C_9^{\rm NP}(\mu_W), C_{10}^{\rm NP})$ plane, representing 
solutions in the four allowed quadrants in this space. However, a high
density scan over all the parameter space shows that the 
allowed solutions in the models considered by us are scattered mostly
around the $(C_9^{\rm NP}(\mu_W), C_{10}^{\rm NP}) =(0,0)$ region, for
the two branches for the quantities $R_{7,8}(\m_W)$, i.e. for 
$R_{7,8}(\m_W)>0$ and $R_{7,8}(\m_W)<0$.  We present the resulting
constraints on the
supersymmetric masses $M_{{\tilde t}_2}$ (mass of the lighter of the two
stop mass eigenstate), $M_{H^\pm}$ (the charged Higgs boson masses) and
$\tan \beta$ (ratio of the two Higgs vacuum expectation values), in the
context of the MFV-MSSM framework, and on the mass insertion
parameter $(\delta_{23})$ in the MIA framework. This updates similar
results worked out along these lines in Ref.~\cite{Hewett:1996ct}.

This paper is organized as follows: In Section \ref{Hamiltonian}, we
enlist the current measurements of the rare $B$-decays which we have
analyzed. The effective Hamiltonian for the SM and the supersymmetric
models studied by us is also given here. In Section \ref{BSLL}, we
present the NNLO implementation of the inclusive and exclusive
$b\to s \ell^+ \ell^-$ transitions that we consider. In Section
\ref{BSSTARLL} we discuss the branching ratios for the exclusive decays
$B \to K^{(*)} \ell^+ \ell^-$ in the SM.  In Section \ref{BSGconstraints},
we study the constraints on the supersymmetric
parameters resulting from the $B\to X_s \g$ decay in the NLO
approximation. In Section \ref{BSLLconstraints}, we present the results
of the model-independent analysis of the $b\to s \ell^+ \ell^-$ modes
based on current data. In Section \ref{MIA}, we describe the specific
SUSY model that we study and present the bounds on the relevant mass
insertions. In Section \ref{Summary}, we summarize our results. Some
loop-functions encountered in the calculation and the stop and chargino
mass matrices are given in the appendices.

\section{Effective Hamiltonian}
\label{Hamiltonian}
The effective Hamiltonian in the SM inducing the $b\to s \ell^+
\ell^-$ and $b\to s\g$ transitions can be expressed as follows:
\begin{eqnarray}
    \label{Heff}
    {\cal H}_{\eff} =  - \frac{4G_F}{\sqrt{2}} V_{ts}^* V_{tb}
    \sum_{i=1}^{10} C_i(\mu) \, O_i (\mu)\quad ,
\end{eqnarray}
where $O_i(\mu)$ are dimension-six operators at the scale $\mu$, 
$C_i(\mu)$ are the corresponding Wilson coefficients, $G_F$ is the
Fermi coupling constant, and the CKM dependence has been made
explicit.  The operators can be chosen as \rf{BMU}
\begin{equation}
\label{oper}
\begin{array}{rclrcl}
    O_1    & = & (\bar{s}_{L}\gamma_{\mu} T^a c_{L })
                (\bar{c}_{L }\gamma^{\mu} T^a b_{L}) \, , &
    O_2    & = & (\bar{s}_{L}\gamma_{\mu}  c_{L })
                (\bar{c}_{L }\gamma^{\mu} b_{L}) \, , \\ \vspace{0.2cm}
    O_3    & = & (\bar{s}_{L}\gamma_{\mu}  b_{L })
                \sum_q (\bar{q}\gamma^{\mu}  q) \, , &
    O_4    & = & (\bar{s}_{L}\gamma_{\mu} T^a b_{L })
                \sum_q (\bar{q}\gamma^{\mu} T^a q) \, , \\ \vspace{0.2cm}
    O_5    & = & (\bar{s}_L \gamma_{\mu_1} \gamma_{\mu_2}
                \gamma_{\mu_3}b_L)
                \sum_q(\bar{q} \gamma^{\mu_1} \gamma^{\mu_2}\gamma^{\mu_3}q)
                \, , &
    O_6    & = & (\bar{s}_L \gamma_{\mu_1} \gamma_{\mu_2}
                \gamma_{\mu_3} T^a b_L)
                \sum_q(\bar{q} \gamma^{\mu_1} \gamma^{\mu_2}
                \gamma^{\mu_3} T^a q) \, , \vspace{0.2cm} \\
\vspace{0.2cm}
    O_7    & = & \frac{e}{g_s^2} m_b (\bar{s}_{L} \sigma^{\mu\nu}
                b_{R}) F_{\mu\nu} \, , &
    O_8    & = & \frac{1}{g_s} m_b (\bar{s}_{L} \sigma^{\mu\nu}
                T^a b_{R}) G_{\mu\nu}^a \, , \\ \vspace{0.2cm}
    O_9    & = & \frac{e^2}{g_s^2}(\bar{s}_L\gamma_{\mu} b_L)
                \sum_\ell(\bar{\ell}\gamma^{\mu}\ell) \, , &
    O_{10} & = & \frac{e^2}{g_s^2}(\bar{s}_L\gamma_{\mu} b_L)
                \sum_\ell(\bar{\ell}\gamma^{\mu} \gamma_{5} \ell) \, ,
\end{array}
\end{equation}
where the subscripts $L$ and $R$ refer to left- and right- handed
components of the fermion fields. We work in the approximation where
the combination $(V_{us}^* V_{ub})$ of the Cabibbo-Kobayashi-Maskawa
(CKM) matrix elements \cite{ckm} is neglected; in this case the CKM
structure factorizes, as indicated in \eq{Heff}. Of course, for the
sake of book keeping, one can keep the individual top-quark and
charm-quark contributions in the loop separately, but as there is no
way to distinguish these individual contributions we will give the
results in the summed form.

Note the inverse powers of $g_s$ in the definition of the operators
$O_7$,...,$O_{10}$ in Eq. (\ref{oper}). These factors have been
introduced by Misiak in Ref. \cite{Misiak93} in order to simplify the
organization of the calculations. In this framework, the LO result for the
$b \to s \ell^+
\ell^-$ decay amplitude is obtained in the following three steps: the
matching conditions $C_i(\mu_W)$ have to be worked out at
$O(\alpha_s^0)$, the renormalization group evolution has to be
performed using the $O(\alpha_s^1)$ anomalous dimension matrix and the
matrix elements of the operators $O_i$ have to be worked out at order
$1/\alpha_s$. In going to the NLO precision all the three steps have
to be improved by one order in $\alpha_s$.

At an arbitrary scale $\mu$ the Wilson coefficients can be decomposed
as
\begin{equation}
\label{decomp}
C_i(\mu) = C_i^{(0)}(\mu) + \frac{\alpha_s(\mu)}{4\pi} \,  C_i^{(1)}(\mu)
 + \frac{\alpha^2_s(\mu)}{(4\pi)^2} \,  C_i^{(2)}(\mu) + \ldots \quad .
\end{equation}
We note that in our basis only $C_2$ is different from zero at the
matching scale $\mu_W$ at leading order, viz. $C_i^{(0)}(\mu_W) =
\delta_{i2}$. At the low scale $\mu_b$ (of order $m_b$), the
coefficients $C_i^0(\mu_b)$ are non-zero for $i=1,...,6;9$ whereas
they vanish for $i=7,8,10$.

We shall use this effective Hamiltonian and calculate the matrix
elements for the decays of interest, specifying the degree of
theoretical accuracy.
     
The experimental input that we use in our analysis is given
below. Except for the inclusive branching ratio for $B\to X_s \g$,
which is the average of the results from CLEO, ALEPH and BELLE
measurements \cite{cleobsg,alephbsg,bellebsg}, all other entries are
taken from the two BELLE papers listed in Ref.~\cite{bellebsll}:
\bea
\branch (B\to X_s \g) &=& (3.22\pm 0.40) \times 10^{-4} \; , 
\label{bsgexp}\\
\branch (B\to K \mu^+ \mu^-) &=& (0.99^{+0.40+0.13}_{-0.32-0.14})
 \times 10^{-6} \; , \label{bkmmexp}\\
\branch (B\to K e^+ e^-) &=& (0.48^{+0.32+0.09}_{-0.24-0.11})\times
10^{-6} \; , \label{bkeeexp} \\
\branch (B\to K \ell^+ \ell^-) &=& (0.75^{+0.25}_{-0.21}\pm 0.09)\times
10^{-6} \; , \label{bkllexp} \\
\branch (B\to K^* \mu^+ \mu^-) &\leq& 3.0 \times  10^{-6} \; {\rm at}\;
 \cl{90}  \; \label{bksmmexp} , \\
\branch (B\to K^* e^+ e^-) &\leq& 5.1 \times  10^{-6} \; {\rm at}\;   
 \cl{90} \;  \label{bkseeexp} ,\\ 
\branch (B\to X_s \mu^+ \mu^-) &\leq & 19.1 \times  10^{-6} \; {\rm at}\;
 \cl{90} \;   \label{bsmmexp} ,\\
\branch (B\to X_s e^+ e^-) &\leq& 10.1 \times  10^{-6} \; {\rm
at}\; \cl{90}
 \;  . \label{bseeexp}
\eea
The experimental numbers given in Eqs.~(\ref{bkmmexp}) --
(\ref{bseeexp}) refer to the so--called non--resonant branching ratios
integrated over the entire dilepton invariant mass spectrum. In the
experimental analyses, judicious cuts are used to remove the dominant
resonant contributions arising from the decays $B \to (X_s,K,K^*)
(J/\psi,\psi^\prime,...) \to (X_s,K,K^*) \ell^+ \ell^-$. A direct
comparison of experiment and theory is, of course, very desirable, but
we do not have access to this restricted experimental
information. Instead, we compare the theoretical predictions with data
which has been corrected for the experimental acceptance using
SM-based theoretical distributions from Ref.~\cite{Ali:1997bm,
Ali:2000mm}. In the present analysis, we are assuming that the
acceptance corrections have been adequately incorporated in the
experimental analysis in providing the branching ratios and upper
limits listed above. We will give the theoretical branching ratios
integrated over all dilepton invariant masses to compare with these
numbers. However, for future analyses, we emphasize the dilepton
invariant mass distribution in the low-$\hat{s}$ region, $\hat{s}
\equiv m_{\ell^+ \ell^-}^2/m_{b,  pole}^2 \leq 0.25$, where the NNLO
calculations for the inclusive decays are known, and resonant effects
due to $J/\psi$, $\psi'$, etc. are expected to be small.

\section{Inclusive $b\to s \ell^+ \ell^-$ decays at NNLO}
\label{BSLL}
We start by discussing the NNLO analysis of the $B\to X_s \ell^+
\ell^-$ decays presented in Refs.~\cite{BMU,AAGW}, recalling that the
$O(\alphas)$ corrections to the matrix elements computed in \rf{AAGW}
have been calculated only below the $c\bar c$ resonances.

In the NNLO approximation, the invariant dilepton mass distribution
for the inclusive decay $B \to X_s \ell^+ \ell^-$ can be written as
\begin{eqnarray}
\label{rarewidth}
    &&\frac{d\Gamma(b\to X_s \ell^+\ell^-)}{d\hat s}=
    \left(\frac{\alpha_{em}}{4\pi}\right)^2
    \frac{G_F^2 m_{b,pole}^5\left|V_{ts}^*V_{tb}^{}\right|^2}
    {48\pi^3}(1-\hat s)^2 \times \nn \\
    &&\left ( \left (1+2\hat s\right)
    \left (\left |\widetilde C_9^{\eff}\right |^2+
    \left |\widetilde C_{10}^{\eff}\right |^2 \right )
    + 4\left(1+2/\hat s\right)\left
    |\widetilde C_7^{\eff}\right |^2+
    12 \mbox{Re}\left (\widetilde C_7^{\eff}
    \widetilde C_9^{\eff*}\right ) \right ) \, .
\end{eqnarray}
In the SM the effective Wilson coefficients $\tilde{C}_7^{\eff}$,
$\tilde{C}_9^{\eff}$ and $\tilde{C}_{10}^{\eff}$ are given
by~\cite{BMU,AAGW}
\begin{eqnarray}
    \widetilde C_7^{\eff}&=&\left (1+\frac{\alpha_s(\mu)}{\pi}
    \omega_7 (\hat{s})\right ) \, A_7  \nonumber \\
    && -\frac{\alpha_{s}(\mu)}{4\pi}\left(C_1^{(0)} F_1^{(7)}(\hat{s})+
    C_2^{(0)} F_2^{(7)}(\hat{s})+A_8^{(0)} F_8^{(7)}(\hat{s})\right) \, ,
    \label{effcoeff7}  \\
    \widetilde C_9^{\eff}&=&\left (1+\frac{\alpha_s(\mu)}{\pi}
    \omega_9 (\hat{s})\right )
    \left (A_9 + T_9 \, h (\hat m_c^2,
    \hat{s})+U_9 \, h (1,\hat{s})+
    W_9 \, h (0,\hat{s})\right) \nonumber \\
    && -\frac{\alpha_{s}(\mu)}{4\pi}\left(C_1^{(0)} F_1^{(9)}(\hat{s})+
    C_2^{(0)} F_2^{(9)}(\hat{s})+
    A_8^{(0)} F_8^{(9)}(\hat{s})\right) \, , \label{effcoeff9} \\
    \widetilde C_{10}^{\eff}&=& \left (1+
    \frac{\alpha_s(\mu)}{\pi}
    \omega_9 (\hat{s})\right ) \, A_{10} \, , \label{effcoeff10} 
\end{eqnarray}
where the functions $h (\hat m_c^2,\hat{s})$ and $\omega_9(\hat{s})$
are given in Ref.~\cite{BMU}, while $\omega_7(\hat{s})$ and
$F_{1,2,8}^{(7,9)}(\hat{s})$ can be seen in Ref.~\cite{AAGW}. The
auxiliary quantities $A_7$, $A_8$, $A_9$, $A_{10}$, $T_9$, $U_9$,
$W_9$ are the following linear combinations of the Wilson coefficients
$C_i(\mu)$ (see Eq.(\ref{Heff})):
\begin{eqnarray}
A_7 &=& \frac{4 \pi}{\alpha_s(\mu)} \, C_7(\mu) 
- \frac{1}{3} \, C_3(\mu)
- \frac{4}{9} \, C_4(\mu)
- \frac{20}{3} \, C_5(\mu)
- \frac{80}{9} \, C_6(\mu) \, ,  \\
A_8 &=& \frac{4 \pi}{\alpha_s(\mu)} \, C_8(\mu) 
+  C_3(\mu)
- \frac{1}{6} \, C_4(\mu)
+ 20 \, C_5(\mu)
- \frac{10}{3} \, C_6(\mu) \, ,  \\
A_9 &=&  \frac{4 \pi}{\alpha_s(\mu)} \, C_9(\mu) + 
\sum_{i=1}^{6} \, C_i(\mu) \, \gamma_{i9}^{(0)} \, \ln \frac{m_b}{\mu}
\nonumber \\ 
&& + \frac{4}{3} \, C_3(\mu)
+ \frac{64}{9} \, C_5(\mu)
+ \frac{64}{27} \, C_6(\mu) \, ,  \\
A_{10} &=&  \frac{4 \pi}{\alpha_s(\mu)} \, C_{10}(\mu) \, , \\
T_9 &=&  
+  \frac{4}{3} \, C_1(\mu) +  C_2(\mu)
+ 6 \, C_3(\mu)
+ 60 \, C_5(\mu) \, ,  \\
U_9 &=&  
- \frac{7}{2} \, C_3(\mu) - \frac{2}{3} \, C_4(\mu)
-38 \, C_5(\mu)
- \frac{32}{3} \, C_6(\mu) \, ,  \\
W_9 &=&  
- \frac{1}{2} \, C_3(\mu) - \frac{2}{3} \, C_4(\mu)
-8 \, C_5(\mu)
- \frac{32}{3} \, C_6(\mu) \, .
\end{eqnarray}
The elements $\gamma_{i9}^{(0)}$ can be seen in Eq. (26) of ref. 
\cite{BMU}. $A_8^{(0)}$ in Eqs. (\ref{effcoeff7}) and  (\ref{effcoeff9})
denotes the lowest order piece of $A_8$: 
\begin{equation}
A_8^{(0)} =  C_8^{(1)}(\mu) 
+  C_3^{(0)}(\mu)
- \frac{1}{6} \, C_4^{(0)}(\mu)
+ 20 \, C_5^{(0)}(\mu)
- \frac{10}{3} \, C_6^{(0)}(\mu) \, . 
\end{equation}
The numerical values for the coefficients $A_7$, $A_8^{(0)}$, $A_9$,
$A_{10}$, $T_9$, $U_9$, $W_9$, $C_1$ and $C_2$ are
obtained after solving the renormalization group equations for the
Wilson coefficients $C_i(\mu)$, using the matching conditions from
ref. \cite{BMU} and the anomalous dimension matrices from
Refs. \cite{BMU,Chetyrkin:1997vx}.  As mentioned earlier, we do not
separate charm- and top- quark contributions and perform the matching
(for both) at the scale $\mu_W=m_W$.  The resulting values are
summarized in Table~\ref{table:coeff}.
Note that when calculating the decay width (\ref{rarewidth}), we
retain only terms linear in $\alpha_s$ (and thus in $\omega_9$ and
$\omega_7$) in $|\widetilde C_9^{\eff}|^2$ and $|\widetilde
C_7^{\eff}|^2$.  In the interference term $\mbox{Re} \left (\widetilde
C_7^{\eff} \widetilde C_9^{\eff*} \right )$ too, we keep only terms
linear in $\alpha_s$. By construction, one has to make the
replacements $\omega_9 \to \omega_{79}$ and $\omega_7\to \omega_{79}$
in this term where the function $\omega_{79}(\hat{s})$ can be found in
Ref.~\cite{AAGW}.

We now turn to the modifications of the effective Wilson coefficients
$\widetilde C_7^{\eff}$, $\widetilde C_9^{\eff}$ and $\widetilde
C_{10}^{\eff}$ in the presence of new physics which enters through a
modification of the Wilson coefficients $C_7$, $C_8$, $C_9$ and
$C_{10}$ at the matching scale $\mu_W$. By doing so, we tacitly assume
that the scale of new physics is close enough to the weak scale $m_W$,
justifying to integrate out simultaneously the heavy SM particles and
the additional ones present in the new physics scenario. For
simplicity we assume that only the lowest non-trivial order of these
Wilson coefficients get modified by new physics, which in our setup
(see Eqs. (\ref{Heff}),(\ref{oper}),(\ref{decomp})) means that
$C_7^{(1)}(\mu_W)$, $C_8^{(1)}(\mu_W)$, $C_9^{(1)}(\mu_W)$,
$C_{10}^{(1)}(\mu_W)$ get modified. The shifts of the Wilson
coefficients at $\mu_W$ can be written as:
\begin{equation} 
C_i(\mu_W) \longrightarrow C_i(\mu_W) + \frac{\alpha_s}{4\pi} \,
C_i^{NP}(\mu_W) \, .
\end{equation} 
These shift at the matching scale are translated through the RGE step
into modifications of the coefficients $C_i (\m_b)$ at the low scale
$\mu_b$, leading in turn to modifications of the effective Wilson
coefficients defined in Eqs.
(\ref{effcoeff7}--\ref{effcoeff10}). They now read
\begin{eqnarray}
    \widetilde C_7^{\eff}&=&\left (1+\frac{\alpha_s(\mu)}{\pi}
    \omega_7 (\hat{s})\right ) ( A_7 + A_{77} \; C_7^{NP}(\mu_W) + A_{78} \; 
     C_8^{NP}(\mu_W) ) \nonumber \\
    && -\frac{\alpha_{s}(\mu)}{4\pi}\left(C_1^{(0)} F_1^{(7)}(\hat{s})+
    C_2^{(0)} F_2^{(7)}(\hat{s})
+A_8^{(0)} F_8^{(7)}(\hat{s})
+A_{88}^{(0)} \, C_8^{NP}(\mu_W) \, F_8^{(7)}(\hat{s})
\right) \, ,
    \label{effcoeff7new}  \\
    \widetilde C_9^{\eff}&=&\left (1+\frac{\alpha_s(\mu)}{\pi}
    \omega_9 (\hat{s})\right )
    \left (A_9 + T_9 \, h (\hat m_c^2,
    \hat{s})+U_9 \, h (1,\hat{s}) +
    W_9 \, h (0,\hat{s})  + C_9^{NP}(\mu_W) \right) \nonumber \\
    && -\frac{\alpha_{s}(\mu)}{4\pi}\left(C_1^{(0)} F_1^{(9)}(\hat{s})+
    C_2^{(0)} F_2^{(9)}(\hat{s})+
    A_8^{(0)} F_8^{(9)}(\hat{s})+
    A_{88}^{(0)} \, C_8^{NP}(\mu_W) \, F_8^{(9)}(\hat{s})\right) \, , 
    \label{effcoeff9new} \\
    \widetilde C_{10}^{\eff}&=& \left (1+
    \frac{\alpha_s(\mu)}{\pi}
    \omega_9 (\hat{s})\right ) (A_{10} + C_{10}^{NP} ) \, . 
   \label{effcoeff10new} 
\end{eqnarray}
The numerical values for the parameters $A_{77}$, $A_{78}$,
$A_{88}^{(0)}$, which incorporate the effects from the running, are
listed in Table~\ref{table:coeff}.
\begin{table}[htb]
\begin{center}
\begin{tabular}{| l | c | c | c |}\hline\hline
&$\mu=2.5$ GeV& $\mu=5$ GeV  & $\mu=10$ GeV\\ \hline
$\alpha_s$ & $0.267$&  $0.215$ & $  0.180$\\ \hline
$(C_1^{(0)},C_1^{(1)})$ & $(-0.697,0.241)$ & $(-0.487,0.207)$ & $(-0.326,0.184)$\\ \hline
$(C_2^{(0)},C_2^{(1)})$ & $(1.046,-0.028) $& $(1.024,-0.017)$ &  $(1.011,-0.010)$\\ \hline
$(A_7^{(0)},A_7^{(1)})$ & $ (-0.353, 0.023)$ & $(-0.312,0.008)$   & $ (-0.278,-0.002)$\\ \hline
$(A_{77}^{(0)},A_{77}^{(1)})$ & $ (0.577, -0.0524)$ & $(0.672,-0.0391)$ & $(0.760,-0.0277)$ \\ \hline
$(A_{78}^{(0)},A_{78}^{(1)})$ & $ (0.109,-0.00520)$ & $(0.0914,-0.00193)$ & $(0.0707,-0.000263)$ \\ \hline
$A_8^{(0)}$ &$ -0.164 $& $ -0.148$& $ -0.134$\\ \hline
$A_{88}^{(0)}$ &$ 0.618 $& $ 0.706$& $ 0.786$\\ \hline
$(A_9^{(0)},A_9^{(1)})$ & $ (4.287,-0.218) $& $(4.174, -0.035)$ & $ (4.177, 0.107)$ \\ \hline
$(T_9^{(0)},T_9^{(1)})$ & $ (0.114, 0.280) $& $ (0.374,0.252)$ & $ (0.575,0.231)$ \\ \hline
$(U_9^{(0)},U_9^{(1)})$ & $ (0.045,0.023)$ & $ (0.033,0.015)$ & $ (0.022,0.010)$ \\ \hline
$(W_{9}^{(0)},W_{9}^{(1)})$ & $(0.044,0.016)$ & $ (0.032,0.012)$ & $ (0.022,0.008)$ \\ \hline
$(A_{10}^{(0)},A_{10}^{(1)})$ & $ (-4.592,0.379)$ &  $(-4.592,0.379)$ &  $ (-4.592,0.379)$ \\
\hline\hline
\end{tabular}
\end{center}
\caption{Coefficients appearing in
Eqs.~(\ref{effcoeff7}--\ref{effcoeff10}) and
Eqs.~(\ref{effcoeff7new}--\ref{effcoeff10new}) for three different
scales $\mu = 2.5$ GeV, $\mu =5$ GeV and $\mu = 10$ GeV. For
$\alpha_s(\mu)$ (in the $\overline{\mbox{MS}}$ scheme) we used the
two-loop expression with 5 flavours and $\a_s(m_Z)=0.119$.  The
entries correspond to the $\overline{MS}$ top quark mass renormalized
at the scale $m_W$, $m_t (m_W)= 175.9\;\gev$. The superscript $(0)$
refers to the lowest order quantities while the superscript $(1)$
denotes the correction terms of order $\alpha_s$, {\it i.e.}
$X=X^{(0)}+X^{(1)}$ with $X=C,A,T,U,W$.}
\label{table:coeff}
\end{table}
\subsection{Power corrections in inclusive $B\to X_s \ell^+ \ell^-$
decays }
\label{BSLLPOW}
Before presenting a theoretical analysis of the available data on rare
$B$-decays, we would like to discuss power corrections in the
inclusive $B\to X_s \ell^+ \ell^-$ decays.  In the NNLO approximation
and including leading order power corrections in $1/m_b$
\cite{Ali:1997bm} and $1/m_c$ \cite{Buchalla:1997ky}, the invariant
dilepton mass distribution for the inclusive decay $B \to X_s \ell^+
\ell^-$ can be written as
\begin{eqnarray}
\label{eq:rarewidthpower}
    \frac{d\Gamma(b\to s \ell^+\ell^-)}{d\hat s} &=&
    \left(\frac{\alpha_{em}}{4\pi}\right)^2
    \frac{G_F^2 m_{b,pole}^5\left|V_{ts}^*V_{tb}^{}\right|^2}
    {48\pi^3}(1-\hat s)^2 
    \left[ \left (1+2\hat s\right)
    \left (\left |\widetilde C_9^{\eff}\right |^2+
    \left |\widetilde C_{10}^{\eff}\right |^2 \right ) G_1( \hat s) 
\right. \nn \\
    &+& \left. 4\left(1+2/\hat s\right)\left
    |\widetilde C_7^{\eff}\right |^2 G_2( \hat s) +
    12 \mbox{Re}\left (\widetilde C_7^{\eff}
    \widetilde C_9^{\eff*}\right ) G_3( \hat s) + G_c(\hat s) \right]  \,
.
\end{eqnarray}
where
\begin{eqnarray}
G_1( \hat s) &=& 1+ \frac{\lambda_1}{2 m_b^2}+ 3
\frac{1-15 \hat s^2 +10 \hat s^3}{(1-\hat s )^2 (1+ 2 \hat s)}
\frac{\lambda_2}{2 m_b^2} \\
G_2( \hat s) &=& 1+ \frac{\lambda_1}{2 m_b^2}-3 \frac{6+ 3 \hat s -5
\hat s^3}{(1-\hat s )^2 (2+\hat s)} \frac{\lambda_2}{2 m_b^2}~, \\
G_3( \hat s) &=& 1+ \frac{\lambda_1}{2 m_b^2}-
\frac{5+ 6 \hat s- 7 \hat s^2}{(1-\hat s )^2} \frac{\lambda_2}{2 m_b^2}~.
\end{eqnarray}
The values of the heavy
quark matrix elements $\lambda_1$ and
$\lambda_2$ that we use in our analysis are given in
Table~\ref{table:inputs}. The term denoted by $G_c$ takes
$1/m_c$ corrections into account.  It is written as
\begin{eqnarray}
\label{eq:fc}
G_c(\hat s)=-\frac{8}{9} (C_2-\frac{C_1}{6})\frac{\lambda_2}{m_c^2} 
\mbox{Re} \left(  F^*(r) \left[\widetilde C_9^{\eff} (2+ \hat s)+
\widetilde C_7^{\eff} \frac{1+6 \hat s- \hat s^2}{\hat s} \right]
 \right)~.
\end{eqnarray}
Since our basis in Eq.~(\ref{oper}) is different from the one often
used in the literature i.e.~$\tilde{O}_1= (\bar{s}_{L}\gamma_{\mu}
b_{L })(\bar{c}_{L }\gamma^{\mu} c_{L})$ and
$O_1=\tilde{O}_1/2-O_2/6$, Eq.~(\ref{eq:fc}) differs superficially
from the one reported in \cite{Buchalla:1997ky}.  The function $F(r)$,
where $r=\hat s/(4 \hat{m}_c^2)$, is given
below~\cite{Buchalla:1997ky}:
\begin{equation}\label{frl1}
F(r)=\frac{3}{2r}\left\{ \begin{array}{ll}
\dis\frac{1}{\sqrt{r(1-r)}}\arctan\sqrt{\frac{r}{1-r}}
   -1 &  \qquad\qquad 0< r < 1~, \\
 \dis\frac{1}{2\sqrt{r(r-1)}}\left(
\ln\frac{1-\sqrt{1-1/r}}{1+\sqrt{1-1/r}}+i\pi\right)-1 &
\qquad\qquad r > 1~. \end{array} \right.
\end{equation}

The impact of power corrections in inclusive decays $B\to X_s \ell^+
\ell^-$ at NLO has been studied in \cite{Ali:1998sf} in the SM.  In
the low dilepton mass region and for $q^2$ not too close to the photon
pole where $O_7$ dominates, the $1/m_b$ effects enhance the rate by
$\sim 1 \%$. In the high--$\hat s$ region they become negative and
decrease the rate by few percent. Their magnitude rises more and more
towards the boundary $q^2 \sim m_b^2$, where the expansion in $1/m_b$
breaks down \cite{Ali:1997bm}.
The $1/m_c$ expansion Eq.~(\ref{eq:fc}) is valid everywhere except
near threshold $\hat s=4 \hat{m}_c^2$, and it also fails at the
charmonium resonances $J/\Psi$ and higher ones like 
$\Psi^{\prime}$.  The
$1/m_c$ corrections decrease the rate below the charm threshold and
enhance it above by few percent.

\begin{figure}[H]
\vskip 0.0truein
\centerline{\epsfysize=1.8in
{\epsffile{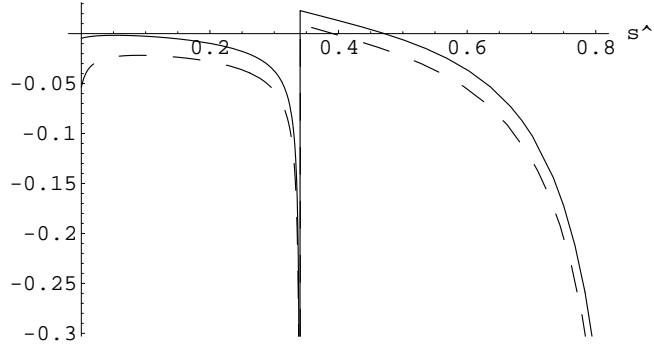}}}
\vskip 0.0truein
\caption{\it Relative size $R(\hat s)$ of the combined $1/m_b$  and
$1/m_c$  power corrections as defined in Eq.~(\ref{eq:size})
in the decay rate in $B\to X_s \ell^+ \ell^-$ decays as a
function of the dilepton invariant mass in the SM (solid)
and for $C_7=-C_7^{SM}$ (dashed).}
\label{fig:mcorr}
\end{figure}

This is illustrated in Figure \ref{fig:mcorr}, where the relative size
$R(\hat s)$ of the combined $1/m_b$ and $1/m_c$ corrections defined as
\begin{eqnarray}
\label{eq:size}
R(\hat s)\equiv \frac{\frac{d\Gamma(b\to X_s \ell^+\ell^-)}{d\hat s}
{\rm (with \;power \, corrections)}-\frac{d\Gamma(b\to X_s
\ell^+\ell^-)}{d\hat s}{\rm (no \; power \,corrections)}}
{\frac{d\Gamma(b\to X_s \ell^+\ell^-)}{d\hat s}{\rm (with \;
power \, corrections)}}
\end{eqnarray}
is shown for the SM, and for comparison also for $C_7=-C_7^{SM}$.
Both $1/m$ correction thus partially cancel in the SM.
The situation with new physics can be different.  In a generic
scenario with $C_7=-C_7^{SM}$ the power corrections can be more
pronounced, in particular for low dilepton mass where both $1/m$
corrections are negative. Together they lower the rate by few percent.
Note that in our estimates of the $B\to X_s \ell^+ \ell^-$ branching
ratio, we include the power corrections in the semileptonic branching
ratios \cite{powerbtoc} as well.
%
%
\subsection{Branching ratios for $B\to X_s \ell^+\ell^-$ in the SM}
In order to eliminate the large uncertainty due to the 
factor $m_{b,pole}^5$ appearing in the decay width for $B \to X_s
\ell^+ \ell^-$, it has become customary to consider instead the
following branching ratio \begin{equation} \branch^{B\to X_s \ell^+
\ell^-} (\hat s)= {\branch_{\hbox{\tiny exp}}^{B \to X_c e\bar{\nu}}
\over \Gamma (B \to X_c e\bar{\nu})} {d\Gamma(B\to X_s \ell^+ \ell^-)
\over d\hat s} \, ,
\end{equation}
in which the factor $m_{b,pole}^5$ drops out. The explicit expression
for the semi-leptonic decay width $\Gamma(B \to X_c e \nu_e)$ can be
found {\it e.g.} in \rf{BMU}. Note that as we are ignoring the
annihilation contributions, which lead to isospin violations in the decay 
widths, and we are using the averaged semileptonic branching ratio to
normalize, all our inclusive branching ratios are to be understood as
averaged over the $B^\pm$ and  $B^0(\overline{B^0})$ decays.

\begin{figure}[H]
\begin{center}
\epsfig{file=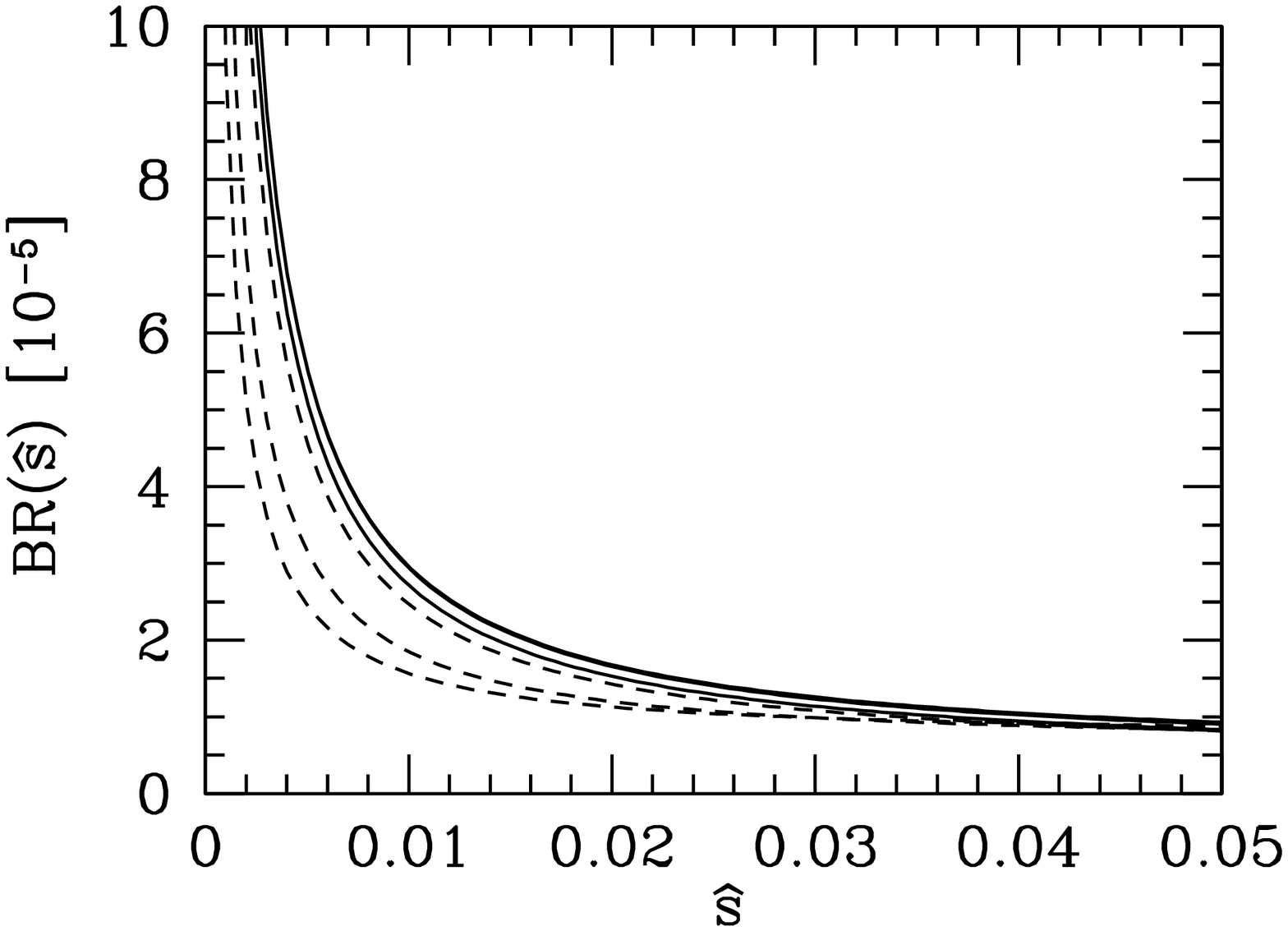,width=0.48\linewidth}
\hspace*{.2cm}
\epsfig{file=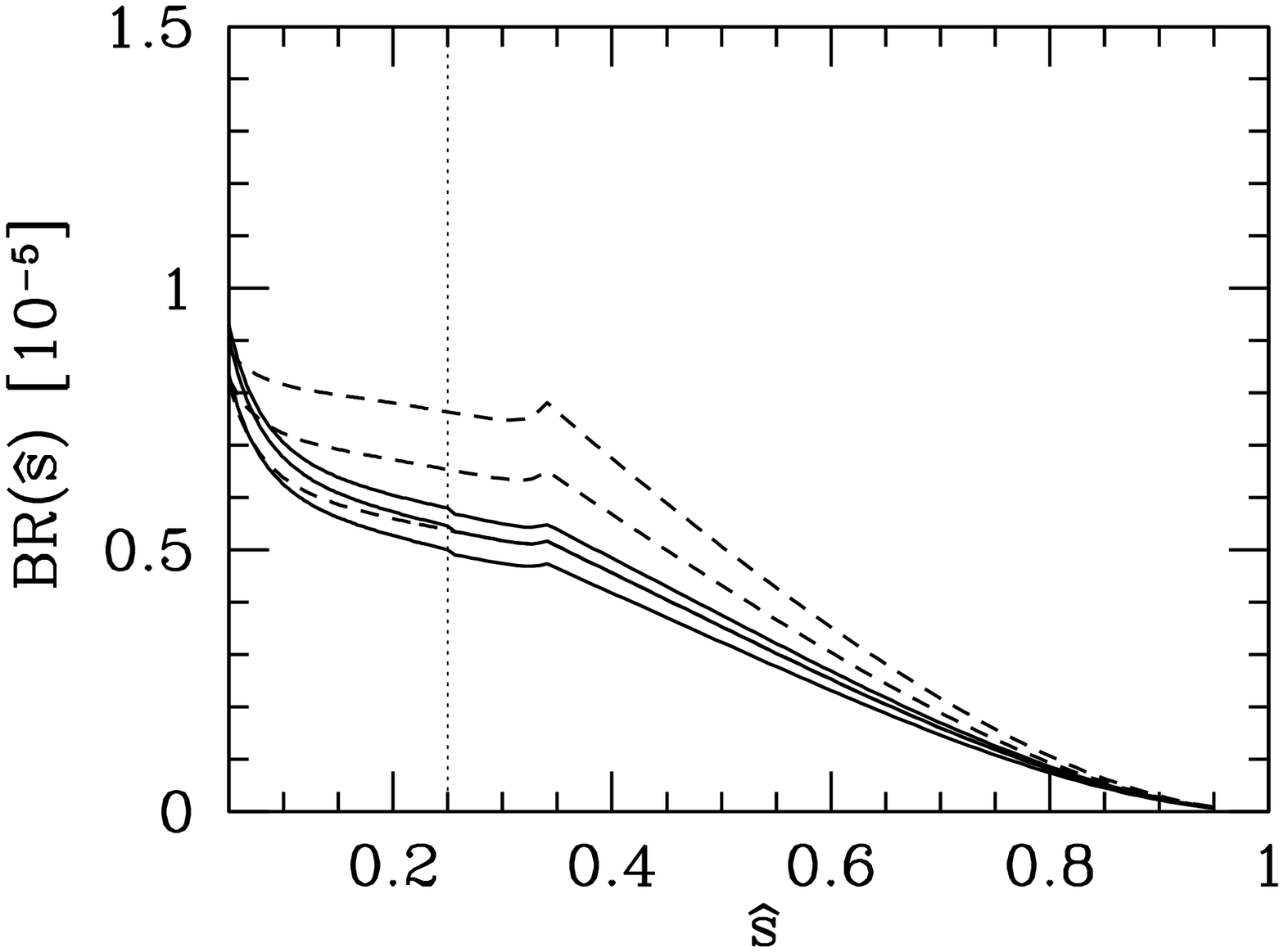,width=0.48\linewidth}
\caption{\it Partial (dashed lines) vs full (solid lines) NNLO
computation of the branching ratio $B\to X_s e^+ e^-$. In the
left plot ($\hat s \in [0,0.05]$) the lowest curves are for
$\m=10\;\gev$ and the uppermost ones for $\mu=2.5\;\gev$. In the right
plot the $\m$ dependence is reversed: the uppermost curves correspond
to $\m=10\;\gev$ and the lowest ones to $\mu=2.5\;\gev$. The right-hand
plot also holds for the decay $B \to X_s \mu^+ \mu^-$.}
\label{fig:nnlo}
\end{center}
\end{figure}

The dilepton invariant mass distribution for the process $B \to X_s
e^+ e^-$ calculated in  NNLO is shown in \fig{fig:nnlo} for
the three choices of the scale $\mu=2.5~\gev$, $\mu=5~\gev$ and
$\mu=~10~\gev$ (solid curves). In this figure, the left-hand plot
shows the distribution in the very low invariant mass region ($\hat s
\in [0,0.05]$, with $0$ to be understood as the kinematic threshold $s =4
m_e^2 \simeq 10^{-6}\; \gev^2$, yielding $\hat{s}=4.5 \times
10^{-8}$), and the right-hand plot shows the dilepton
spectrum in the region beyond $\hat {s} > 0.05$, and hence this also
holds for the decay $B \to X_s \mu^+ \mu^-$. We should stress at this
point that a genuine NNLO calculation only exists for values 
of $\hat{s}$ below 0.25, which is indicated in the right-hand plot 
by the vertical dotted line. For higher values of $\hat{s}$, an
estimate of the NNLO result is obtained by an extrapolation procedure
discussed in more detail at the end of this paragraph.
The so-called partial NNLO dilepton spectrum, 
obtained by switching off
the quantities $F_{1,2,8}^{(7,9)}$ in Eqs. (\ref{effcoeff7}) and 
(\ref{effcoeff9}),
is also shown in each of these cases 
for the same three choices of the scale $\mu$ (dashed curves).
Note that in the left-hand plot, the lowest lying curves are for
$\mu=10~\gev$ and the uppermost ones are for $\mu=2.5\; \gev$. In the
right-hand plot, the scale-dependence is reversed, i.e., the highest
lying curves are for $\mu=~10~\gev$ and the lowest for $\mu=2.5~\gev$.
The crossing (in the partial NNLO BR) happens near $\hat s=0.04$ and
this feature leads to a certain cancellation of the $\mu$ dependence
in the decay rate for $B \to X_s e^+ e^-$. We also note that the NNLO
dilepton invariant mass spectrum in the right-hand plot ($\hat {s} >
0.05$) lies below its partial NNLO counterpart, and hence the partial
branching ratios for both the $B \to X_s e^+ e^-$ and $B \to X_s \mu^+
\mu^-$ decays are reduced in the full NNLO accuracy. More importantly,
from the point of view of our subsequent analysis, \fig{fig:nnlo}
shows that the full NNLO invariant mass distribution is very well
approximated by the partial NNLO for the choice of the scale $\mu=2.5
\;\gev$, in the entire low-$\hat s$ range. This is yet another
illustration of the situation often met in perturbation theory that a
judicious choice of the scale reduces the higher order
corrections. From this observation, it seems reasonable to use
the partial NNLO
curve corresponding to $\mu_b = 2.5 \;
\gev$ as an estimate for the central value of the full NNLO for
$\hat{s}>0.25$.
We estimate the scale dependence in this region by assuming that it is
given by the genuine NNLO calculation at $\hat{s}=0.25$. 

In order to complete our discussion of the computation of the
inclusive branching ratios, it is necessary to discuss their dependence
on the quark masses $m_t$ and $m_c$ (in particular, the latter is
marked for what concerns the SM prediction). We vary both masses
within the errors that we quote in Table~\ref{table:inputs} and
present the results for the branching ratios $B\to X_s e^+ e^-$ and
$B\to X_s \m^+ \m^-$ in Table~\ref{table:smnumbers} where we include
also the power corrections discussed in Sec.~\ref{BSLLPOW}. In
Table~\ref{table:sm} we show the SM central values and the parametric
uncertainties by means of independent error bars (to be interpreted as
$\cl{68}$ uncertainties). In this table, the first error on the
exclusive channel is due to the form factors, and is by far the
dominant one. The other errors in both the exclusive and inclusive
decays come from the scale ($\m_b$), $m_{t,pole}$ and $m_c/m_b$
respectively. Summing the errors in quadrature we get for the
inclusive decays:
\bea
\branch (B\to X_s e^+ e^-) &=& (6.89 \pm 1.01) \times 10^{-6} \;\;\; 
        (\d \branch_{X_see} = \pm 15 \%) \; , \\
\branch (B\to X_s \m^+ \m^-) &=& (4.15 \pm 0.70) \times 10^{-6} \;\;\; 
        (\d \branch_{X_s\m\m} = \pm 17 \%) \; .
\eea
Using the same input parameters, but restricting to the NLO precision,
the inclusive branching ratios have the central values
$\branch (B\to X_s e^+ e^-) = 7.8 \times 10^{-6}$ and 
$\branch (B\to X_s \m^+ \m^-)=5.2 \times 10^{-6}$. Thus, NNLO corrections
reduce the branching ratios by typically $12\%$ and $20\%$, respectively. 
In \rf{Gambino:2001ew}, it was recently suggested in the context
of the decay $B \to X_s \gamma$, where the charm quark mass enters
the matrix elements at the two-loop level only, that it would be
more appropriate to use the running charm mass evaluated at the $\m_b
\simeq O(m_b)$ scale, leading to $m_c/m_b \simeq 0.22$. 
Intuitively, this is a reasonable
choice since the charm quark enters only as virtual particle running
inside loops; formally, on the other hand, it is  also clear that 
the difference between the results obtained
by interpreting $m_c$ as the pole mass or the running mass is a higher
order QCD effect. In what concerns $B \to X_s \ell^+ \ell^-$, the situation
is somewhat different, as the charm quark mass enters in this case also
in the one-loop matrix elements associated with $O_1$ and $O_2$. In these
one-loop contributions, $m_c$ has the meaning of the pole mass
when using the expressions derived in Ref. \cite{AAGW}.
Concerning the charm quark mass in the two-loop expressions, 
the definition $m_c$ is not fixed, like in $B \to X_s \gamma$.
In our analysis, we prefer not to include this effect related to the
definition of the charm quark mass in the final errors that we have
listed.

\section{Exclusive $B\to K^{(*)} \ell^+ \ell^-$ decays}
\label{BSSTARLL}
For what concerns the exclusive decays $B\to K^{(*)} \ell^+ \ell^-$,
we implement the NNLO corrections calculated by Bobeth et al.~in \rf{BMU}
and by Asatrian et al.~in \rf{AAGW} for the short-distance
contribution.  Then, we use the form
factors calculated with the help of the QCD sum rules in
\rf{Ali:2000mm}. Note that, in this case, we have dropped the
contribution to the matrix elements given by the functions $\omega_i
(\hat s)$ since this can be regarded as included in the full QCD form
factors. In adopting this procedure, we are ignoring the so-called
hard spectator corrections, calculated in the decays $B \to K^* \ell^+
\ell^-$ \cite{Beneke:2001at} in the large energy limit of QCD
\cite{Charles:1999gy}, necessarily limiting the invariant mass to the
small-$s$ region. The findings of Ref.~\cite{Beneke:2001at} are that
the dilepton invariant mass distribution in this region is rather
stable against the explicit $O(\alpha_s)$ corrections, and the
theoretical uncertainties are dominated by the form factors
and other non-perturbative parameters specific to the large-energy
factorization approach. This is so, even if one takes the point of
view that the form factor $\xi_\perp(0)$, governing the transition $B
\to K^*$ to the transversely polarized $K^*$-meson, can be assumed
known from the analysis of the radiative transition $B \to K^* \gamma$
in this approach and current data, as it is the contribution of the
longitudinally polarized $K^*$ which dominates the decay rate in the
small-$\hat{s}$ range, for which a knowledge of $\xi_\parallel$ is
required. In principle, using estimates of SU(3)-breaking and HQET,
the function $\xi_\parallel$ for the decays $B \to K^* \ell^+ \ell^-$
can be obtained from the semileptonic decays $B \to \rho \ell
\nu_\ell$. However, as present data on the $Q^2$-dependence in the
decay $B \to \rho \ell \nu_\ell$ is rather sparse and a helicity-based
analysis of the decays $B \to \rho \ell\nu_\ell$ has yet to be
undertaken, one will have to resort to form factor models for
$\xi_\parallel$, which as opposed to the transverse form factor
$\xi_\perp$, is essentially unbounded. In view of this, we ignore the
hard spectator correction and discuss a plausible range of the form
factors in the decays $B \to (K,K^*) \ell^+ \ell^-$.

As already stated, some inference about the magnetic moment form
factor $T_1(0)$, involving the matrix element of the operator $O_7$ in
the decay $B \to K^* \gamma$, has been derived by comparing the
explicit $O(\alpha_s)$ and $\Lambda_{\rm QCD}/M_B$ corrected branching
ratios in the factorization approach with data
\cite{Ali:2001ez,Beneke:2001at,Bosch:2001gv}. One finds that present
data on $B \to K^* \gamma$ decay yields typically a value in the range
$T_1(0)=0.28 \pm 0.04$. This suggests that, including the explicit
$O(\alpha_s)$ corrections, data requires a value of this form factor
which is smaller than its typical QCD sum rule estimate. To
accommodate this, we use the minimum allowed form factors obtained in
the light-cone QCD sum rule formalism, given in Table~5 of
\rf{Ali:2000mm}, as our default set. This, for example, corresponds to
setting $T_1(0)=0.33$. In our numerical analysis, we add a flat $\pm
15\%$ error as residual uncertainty on the form factors. Thus, the
input range for $T_1(0)$ in our analysis $T_1(0)=0.33 \pm 0.05$
overlaps with the phenomenologically extracted value in the
factorization approach given earlier. Again, following the argument
given earlier for the inclusive decays, we set $\mu_b = 2.5 \; \gev$,
and include the NNLO corrections in an analogous fashion to the
inclusive $B \to X_s \ell^+ \ell^-$ case. The explicit expressions for
the $B\to K^{(*)} \ell^+ \ell^-$ branching ratios can be found, for
example, in \rf{Ali:2000mm}.

The input parameters that we use in the analyses are summarized in
Table~\ref{table:inputs}. Our SM predictions for the above discussed
inclusive and exclusive branching ratios are summarized in
Table~\ref{table:sm}. Note that the dominant source of uncertainty
comes from the form factors dependence. Summing the errors in
quadrature we obtain:
\bea
\branch (B\to K \ell^+ \ell^-) &=& (0.35 \pm 0.12) \times 10^{-6} \;\;\; 
	(\d \branch_{K\ell\ell} = \pm 34 \%) \; , \\
\branch (B\to K^* e^+ e^-) &=& (1.58 \pm 0.49) \times 10^{-6} \;\;\; 
	(\d \branch_{K^*ee} = \pm 31 \%) \; , \\
\branch (B\to K^* \m^+ \m^-) &=& (1.19 \pm 0.39) \times 10^{-6} \;\;\; 
	(\d \branch_{K^*\m\m} = \pm 33 \%) \; .
\eea
Note that the dependence of the exclusive decay branching ratios on 
$m_c/m_b$ is much milder, as we are using the
($m_c/m_b$-independent) lifetime $\tau(B^0)$ in
calculating the branching ratios for exclusive decays, as opposed to
the inclusive decays $B \to X_s \ell^+ \ell^-$, where the semileptonic
branching ratios are used for normalization. Since the semileptonic decay
widths depend on $m_c/m_b$, this sensitivity goes over to the inclusive
decay branching ratios for $B \to X_s \ell^+ \ell^-$.
Note also that as we have used $\tau(B^0)$ in
calculating the branching ratios for exclusive decays, all the
branching ratios given above are for the $B^0(\overline{B^0})$ decays.
The ones for the $B^\pm$-decays can be scaled by taking into account
the lifetime difference.

\begin{table}[htb]
\begin{center}
\begin{tabular}{| l | l || l | l |}\hline\hline
$m_Z$        & $91.1867 \; \gev$       & $\a_s (m_Z)$      & 0.119     \gh \\ \hline
$m_W$        & $80.41 \;  \gev$        & $\a_e $           & 1/133     \gh \\ \hline
$m_{b,pole}$ & $4.8 \; \gev$           & $\sin^2 \theta_W$ & 0.23124   \gh \\ \hline
$m_{t,pole}$ & $(173.8 \pm 5)\; \gev$  & $G_F$             &
$1.16639 \times 10^{-5} \; \gev^{-2}$ \gh \\ \hline \hline
$\t_{B^0}$   & $1.54 \; {\rm ps}$ & $|V_{tb}^{}V_{ts}^*|$  & 0.038  \gh \\ \hline
$\branch_{\hbox{\tiny exp}}^{B \to X_c e\bar{\nu}}$ & $0.104$ &
$\l$ &
 $0.225\; $  \gh \\ \hline
$m_c/m_b$ & $0.29\pm 0.04$ & $|V_{tb}|^2 \, |V_{ts}|^2/|V_{cb}|^2$ &
 $0.95$   \gh \\ \hline
$\lambda_1$ & -$0.2~\gev^2$ & $\lambda_2$ & $+0.12~\gev^2$ \gh \\
\hline\hline
\end{tabular}
\end{center}
\caption{Input parameters and their assumed errors used in calculating
the $b\to s \ell^+ \ell^-$ decay rates. The quantities $\lambda$,
$\lambda_1$ and $\lambda_2$ are, respectively, the Wolfenstein
parameter and the two HQET parameters appearing in the heavy quark
expansion.}
\label{table:inputs}
\end{table}
\begin{table}[htb]
\begin{center}

$\branch (B\to X_s e^+ e^-) \times 10^{-6}$

\vskip 0.1cm

\begin{tabular}{| c | c | c | c | c |}\hline\hline
$m_t(\gev)$ & $m_c/m_b$	& $\m_b = 2.5\;\gev$ & $\m_b = 5\;\gev$ & $\m_b =
10\;\gev$ \\ \hline\hline
168.8 & 0.29 & 6.30 & 6.83 & 7.00 \\ \hline
173.8 & 0.29 & 6.52 & 7.08 & 7.26 \\ \hline
178.8 & 0.29 & 6.75 & 7.32 & 7.52 \\ \hline\hline
173.8 & 0.25 & 5.83 & 6.30 & 6.47 \\ \hline
173.8 & 0.29 & 6.52 & 7.08 & 7.26 \\ \hline
173.8 & 0.33 & 7.38 & 8.12 & 8.35 \\ \hline
\end{tabular}

\vskip 0.3cm

$\branch (B\to X_s \m^+ \m^-) \times 10^{-6}$

\vskip 0.1cm

\begin{tabular}{| c | c | c | c | c |}\hline\hline
$m_t(\gev)$ & $m_c/m_b$	& $\m_b = 2.5\;\gev$ & $\m_b = 5\;\gev$ &
$\m_b =
10\;\gev$ \\ \hline\hline 
168.8 & 0.29 & 3.70 & 4.03 & 4.21 \\ \hline
173.8 & 0.29 & 3.88 & 4.23 & 4.42 \\ \hline
178.8 & 0.29 & 4.08 & 4.44 & 4.64 \\ \hline\hline
173.8 & 0.25 & 3.35 & 3.70 & 3.92 \\ \hline
173.8 & 0.29 & 3.88 & 4.23 & 4.42 \\ \hline
173.8 & 0.33 & 4.53 & 4.93 & 5.15 \\ \hline
\end{tabular}
\end{center}
\caption{Dependence of the
inclusive branching ratios $B\to X_s \ell^+ \ell^- \; (\ell=e,\m)$, in
the SM on the scale $\mu_b$, $m_t$ and $m_c/m_b$.}
\label{table:smnumbers}
\end{table}
\begin{table}[htb]
\begin{center}
\begin{tabular}{| l | l |}\hline\hline
$B\to K\ell^+\ell^-$&$\left(0.35\pm 0.11\pm 0.04\pm 0.02\pm 0.0005
 \right) \times 10^{-6}$ \gh \\ \hline
$B\to K^*e^+e^-$&$\left(1.58\pm 0.47\pm 0.12^{+0.06}_{-0.08}\pm 0.04
 \right) \times 10^{-6}$ \gh \\ \hline
$B\to K^*\m^+\m^-$&$\left(1.19\pm 0.36\pm 0.12^{+0.06}_{-0.08}\pm 0.04
 \right) \times 10^{-6}$ \gh \\ \hline
$B\to X_s \mu^+ \mu^-$ & $\left(4.15 \pm 0.27\pm 0.21\pm 0.62 \right)
 \times 10^{-6}$ \gh \\ \hline
$B\to X_s e^+ e^-$ & $\left(6.89 \pm 0.37 \pm 0.25\pm 0.91\right)
 \times 10^{-6}$ \gh \\ \hline\hline
\end{tabular}
\end{center}
\caption{SM predictions at NNLO accuracy for the various inclusive and
exclusive decays involving the quark transition $b\to s \ell^+
\ell^-$. For the exclusive channels the indicated errors correspond to
variations of the form factors, $\mu_b$, $m_{t,pole}$ and $m_c/m_b$,
respectively. For the inclusive channels the errors correspond,
respectively, to variations of $\mu_b$, $m_{t,pole}$ and $m_c/m_b$.}
\label{table:sm}
\end{table}
\section{Model independent constraints from $B\to X_s \g$}
\label{BSGconstraints}
\begin{figure}[H]
\begin{center}
\epsfig{file=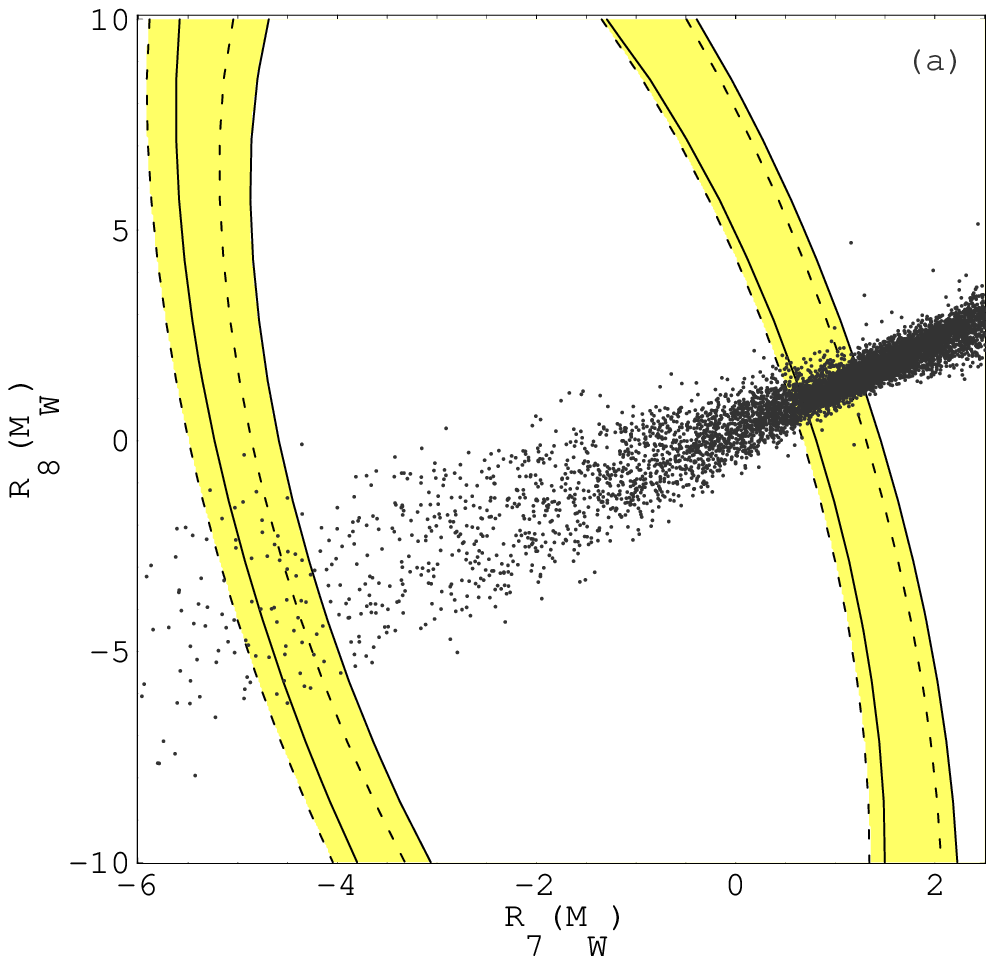,width=0.48\linewidth}
\hspace*{.2cm}
\epsfig{file=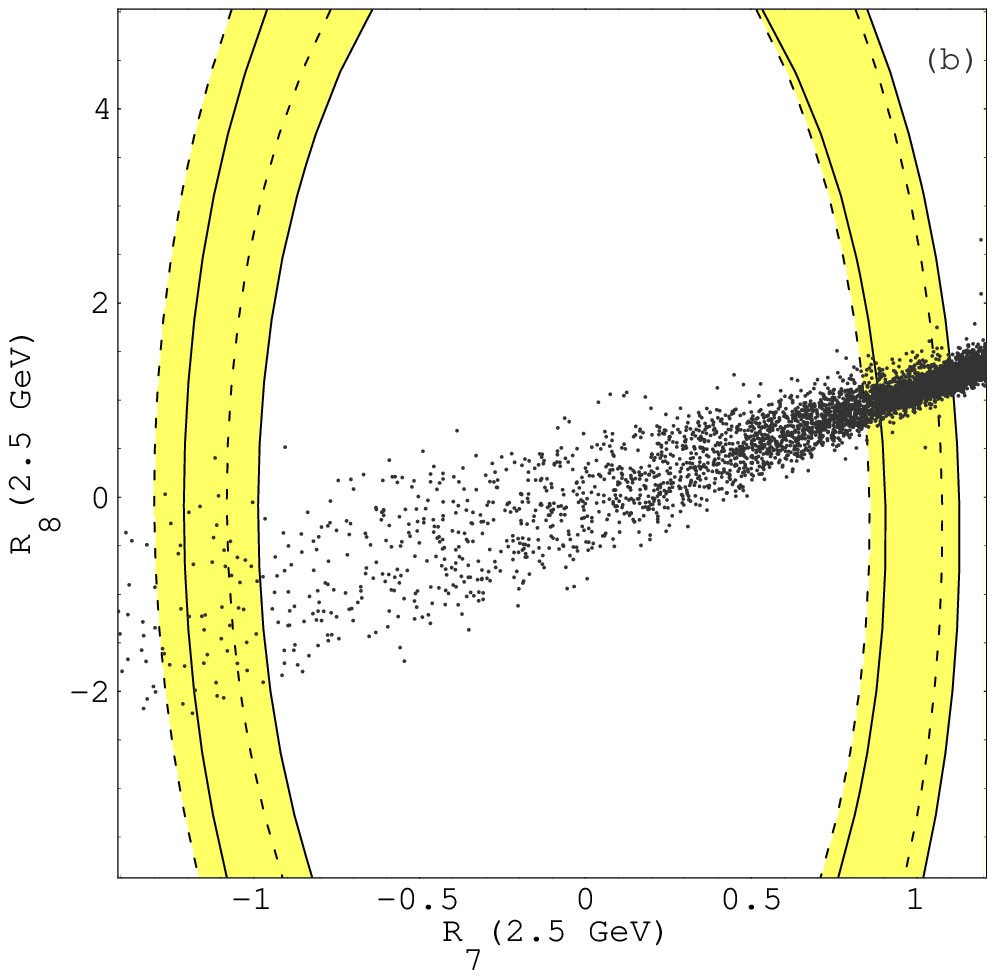,width=0.48\linewidth}
\caption{\it $\cl{90}$ bounds in the $[R_7 (\m), R_8(\m)]$ plane
following from the world average $B\to X_s \g$ branching ratio for
$\mu=m_W$ (left-hand plot) and $\mu=2.5$ GeV (right-hand plot).
Theoretical uncertainties are taken into account. The solid and dashed
lines correspond to the $m_c = m_{c,pole}$ and $m_c =
m_{c}^{\overline{MS}} (\m_b)$ cases respectively. The scatter points
correspond to the expectation in MFV models (the ranges of the SUSY
parameters are specified in the text).}
\label{bsg}
\end{center}
\end{figure}

In this section we work out the $\cl{90}$ bounds that the measurement
(\ref{bsgexp}) implies for $A_7^{\rm tot} (2.5 \; \gev)$, where this
quantity is defined as follows:
\begin{equation}
A_7^{\rm tot} (2.5 \; \gev)  \equiv
A_{77} (2.5 \; \gev) \; C_7^{NP}(\mu_W) + A_{78} (2.5 \; \gev)\;
C_8^{NP}(\mu_W) + A_7^{\rm SM} (2.5 \; \gev)~.
\label{A7tot}
\end{equation}
It was recently pointed out in \rf{Gambino:2001ew} that the charm mass
dependence of the $B\to X_s \g$ branching ratio was underestimate in
all the previous analyses. Indeed, the replacement of the pole mass
($m_{c,pole}/m_{b,pole} = 0.29 \pm 0.02$) with the $\overline{MS}$
running mass ($m_{c}^{\overline{MS}} (\m_b)/m_{b,pole} = 0.22 \pm
0.04$) increases the branching ratio of about $11 \%$. In order to
take into account this additional source of uncertainty, we work out
the constraints on the Wilson coefficients for both choices of the
charm mass; we will then use the loosest bounds in the $b\to s \ell^+
\ell^-$ analysis. We use the numerical expression for the integrated
$B\to X_s \g$ branching ratio as a function of $R_{7,8}(\mu_W)\equiv
C_{7,8}^{\rm tot} (\mu_W) / C_{7,8}^{\rm SM} (\mu_W)$ presented in
\rf{Kagan:1999ym} (Note that, for $m_c/m_b = 0.22$, we had to compute
the small corrections to the coefficients $B_{ij}$). For the sake of
definiteness we shall take $\m_W=m_W$ in deriving the constraints on
physics beyond the SM. We impose the bound (\ref{bsgexp}) at $\cl{90}$
and include the theoretical uncertainty due to the variation of the
scale $\mu_b$ in the range $[m_b/2, 2 m_b]$. In \fig{bsg}a, we present
the resulting allowed regions in the $[R_7(\mu_W),R_8(\mu_W)]$ plane;
the solid and dashed lines correspond to the $m_c = m_{c,pole}$ and
$m_c = m_{c}^{\overline{MS}} (\m_b)$ cases respectively. According to
the analysis presented in \rf{GL}, we restrict, in \fig{bsg}a, to
$|R_8(\mu_W)| \leq 10$ in order to satisfy the constraints from the
decays $B\to X_s g$ and $B\to X_{c \!\! /}$ (where $X_{c\!\! /}$
denotes any hadronic charmless final state). Evolving the allowed
regions to the scale $\mu_b=2.5 \; \gev$ and assuming that new physics
only enters in $C_{7,8}^{(1)}$, we plot in \fig{bsg}b the
corresponding low--scale bounds in the plane $[R_7(2.5 \;
\gev),R_8(2.5 \; \gev)]$, where $R_{7,8}(\mu_b)\equiv A_{7,8}^{\rm
tot} (\mu_b) / A_{7,8}^{\rm SM} (\mu_b)$.  The regions in \fig{bsg}b
translate in the following allowed constraints:
\bea
\cases{
m_c/m_b = 0.29:  \;\;\; A_7^{\rm tot} (2.5 \; \gev)\in [-0.37,-0.18] \; \& \; [0.21,0.40] \; , & \cr 
m_c/m_b = 0.22:  \;\;\; A_7^{\rm tot} (2.5 \; \gev)\in [-0.35,-0.17] \; \& \; [0.25,0.43] \; . & \cr}
\eea
In the subsequent numerical analysis we impose the union of the above
allowed ranges
\bea
\label{a7lim}
 -0.37 \leq A_7^{\rm tot,<0} (2.5 \; \gev) \leq -0.17 
  & \& &
  0.21 \leq A_7^{\rm tot,>0} (2.5 \; \gev) \leq 0.43 
\eea
calling them $A_7^{\rm tot}$--positive and $A_7^{\rm tot}$--negative
solutions.

\section{Model independent constraints from $b\to s \ell^+ \ell^-$} 
\label{BSLLconstraints} 
\begin{figure}[H]
\begin{center}
\epsfig{file=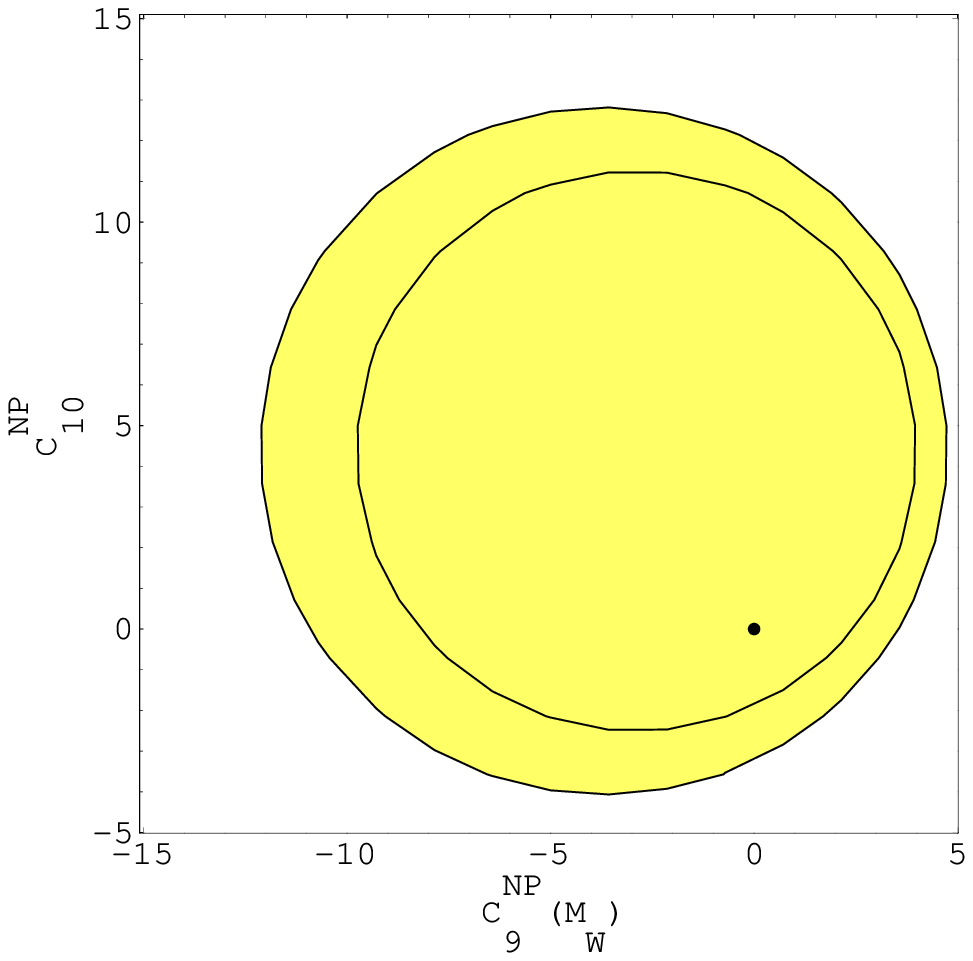,width=0.4\linewidth}
\hskip 0.5cm 
\epsfig{file=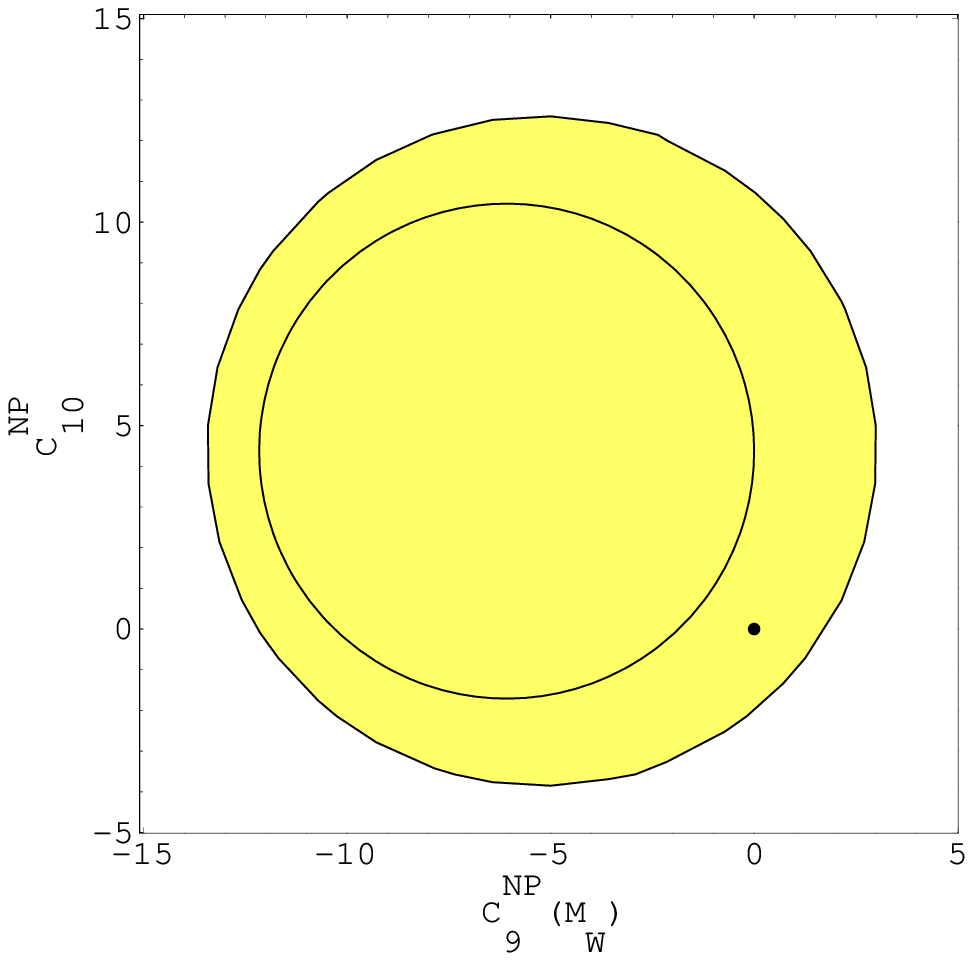,width=0.4\linewidth}
\caption{\it {\bf NLO Case.} Constraints in the
$[C_9^{NP}(\mu_W),C_{10}^{NP}]$ plane that come from the BELLE
$\cl{90}$ upper limit $\branch (B\to X_s e^+ e^-) \leq 10.1 \times
10^{-6}$. Theoretical uncertainties are taken into account.  The plots
correspond to the $A_7^{\rm tot}(2.5 \; \gev)<0$ and $A_7^{\rm
tot}(2.5 \; \gev) >0$ case, respectively. In each plot the outer
contour corresponds to the smaller $|A_7^{\rm tot}|$ value. The dot in
plot on the left is the SM point.}
\label{fig:seeNLO}
\end{center}
\end{figure}

\begin{figure}[H]
\begin{center}
\epsfig{file=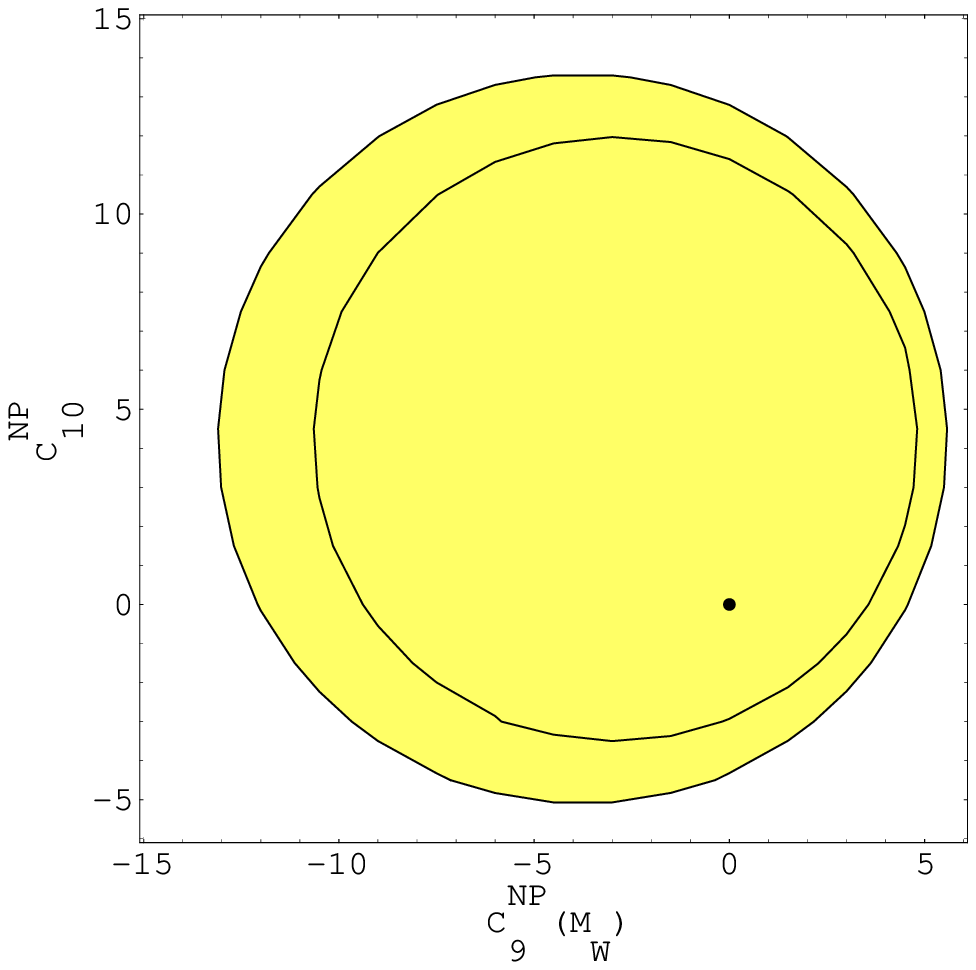,width=0.4\linewidth}
\hskip 0.5cm 
\epsfig{file=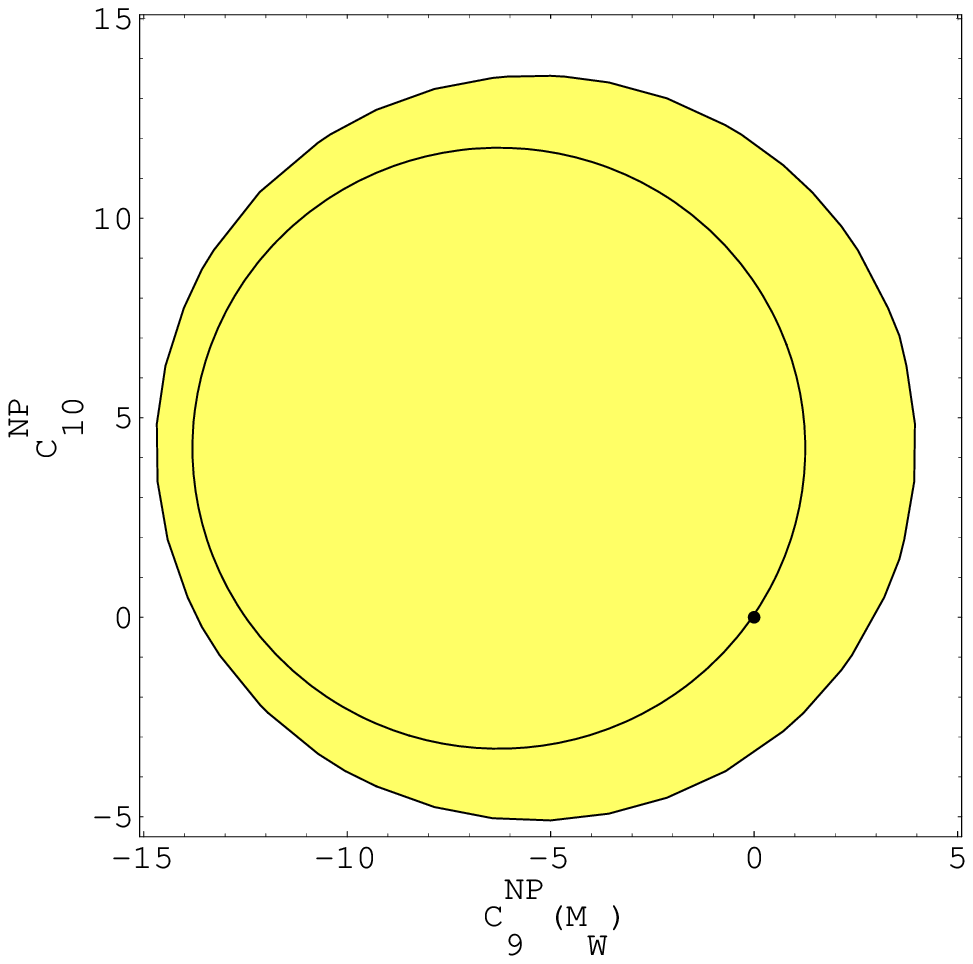,width=0.4\linewidth}
\caption{\it {\bf NNLO Case.} Constraints in the
$[C_9^{NP}(\mu_W),C_{10}^{NP}]$ plane that come from the BELLE
$\cl{90}$ upper limit $\branch (B\to X_s e^+ e^-) \leq 10.1 \times
10^{-6}$.  See \fig{fig:seeNLO} for further details.}
\label{fig:seeNNLO}
\end{center}
\end{figure}

\begin{figure}[H]
\begin{center}
\epsfig{file=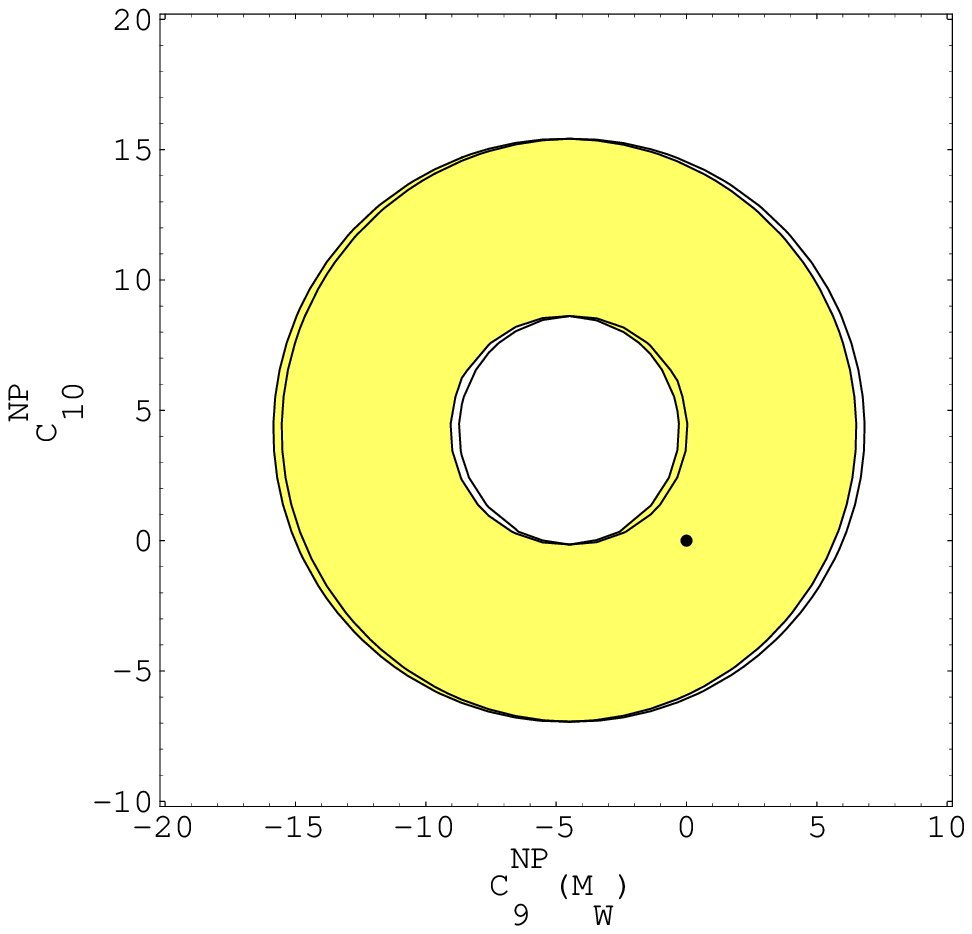,width=0.4\linewidth}
\hskip 0.5cm 
\epsfig{file=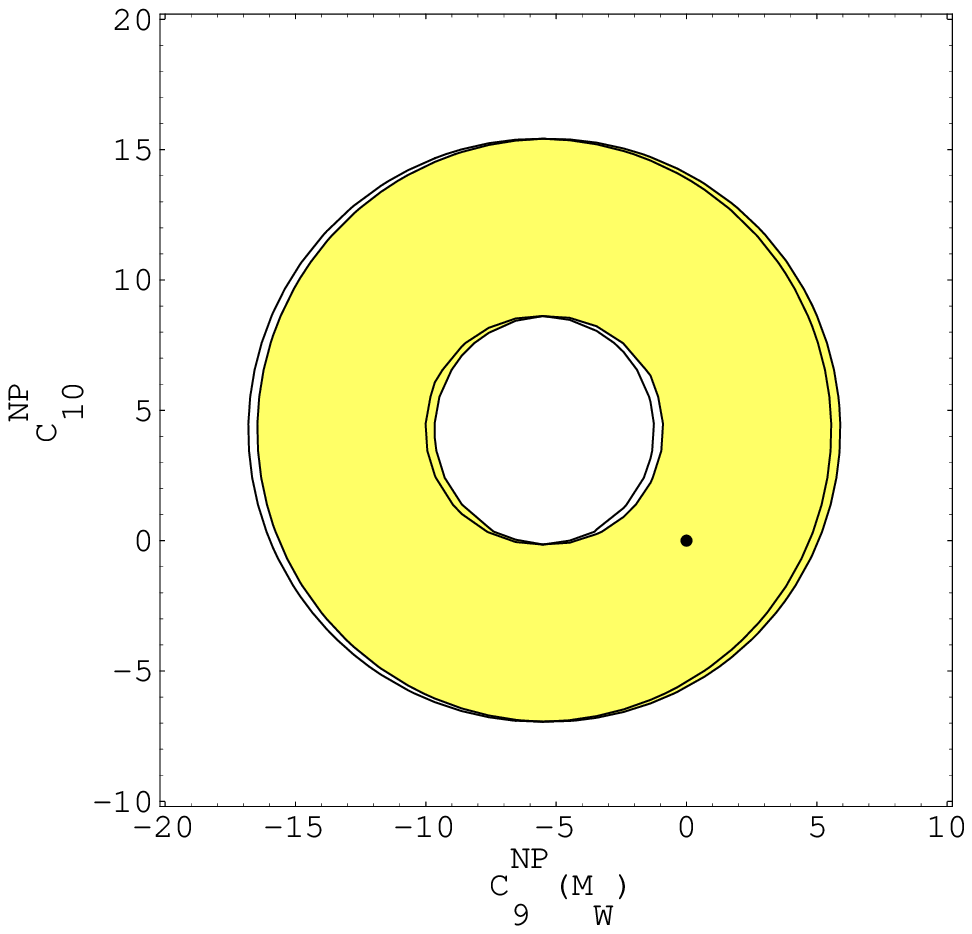,width=0.4\linewidth}
\caption{\it {\bf NNLO Case.} Constraints in the
$[C_9^{NP}(\mu_W),C_{10}^{NP}]$ plane that come from the $90\% CL$
BELLE constraints $0.38 \times 10^{-6} \leq \branch (B\to K \ell^+ \ell^-)
\leq 1.2 \times 10^{-6}$ (Eq.~\ref{bkllexp}). Note that only the 
annular regions between the two circles is allowed.
  See \fig{fig:seeNLO} for further details.}
\label{fig:kllNNLO} 
\end{center}
\end{figure}

\begin{figure}[H]
\begin{center}
\epsfig{file=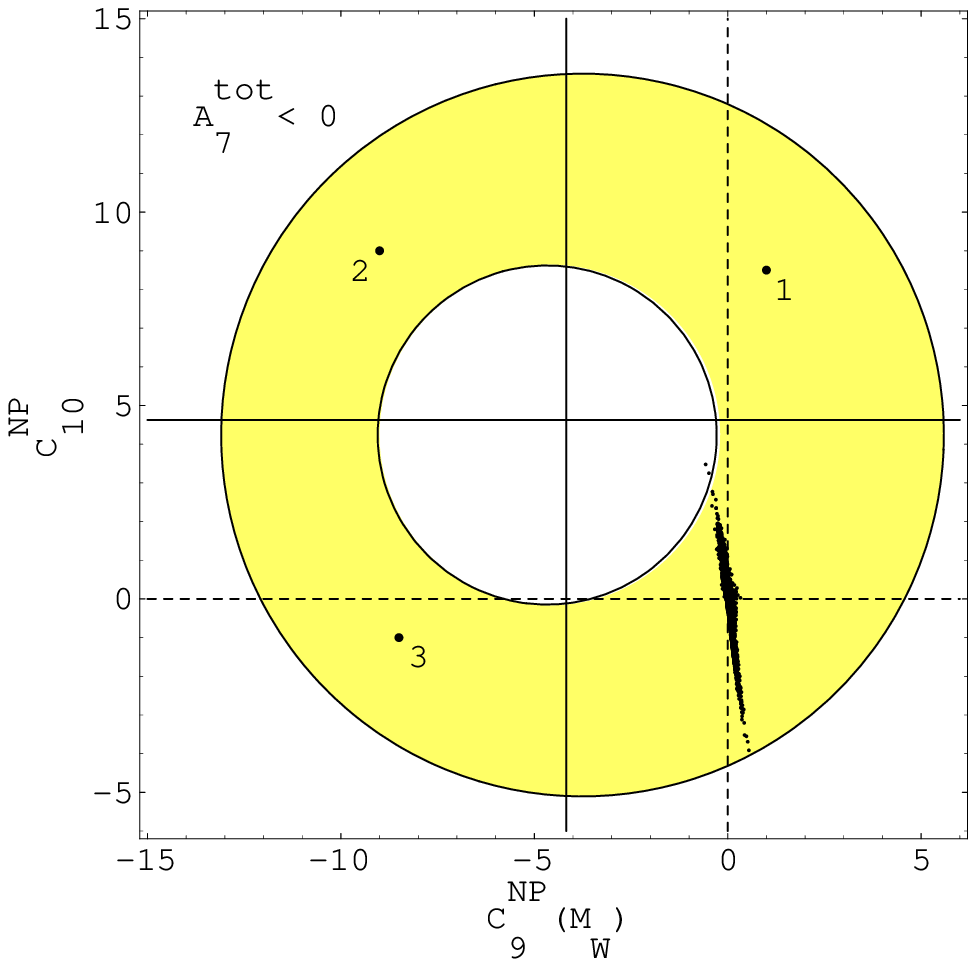,width=0.4\linewidth}
\hskip 0.5cm 
\epsfig{file=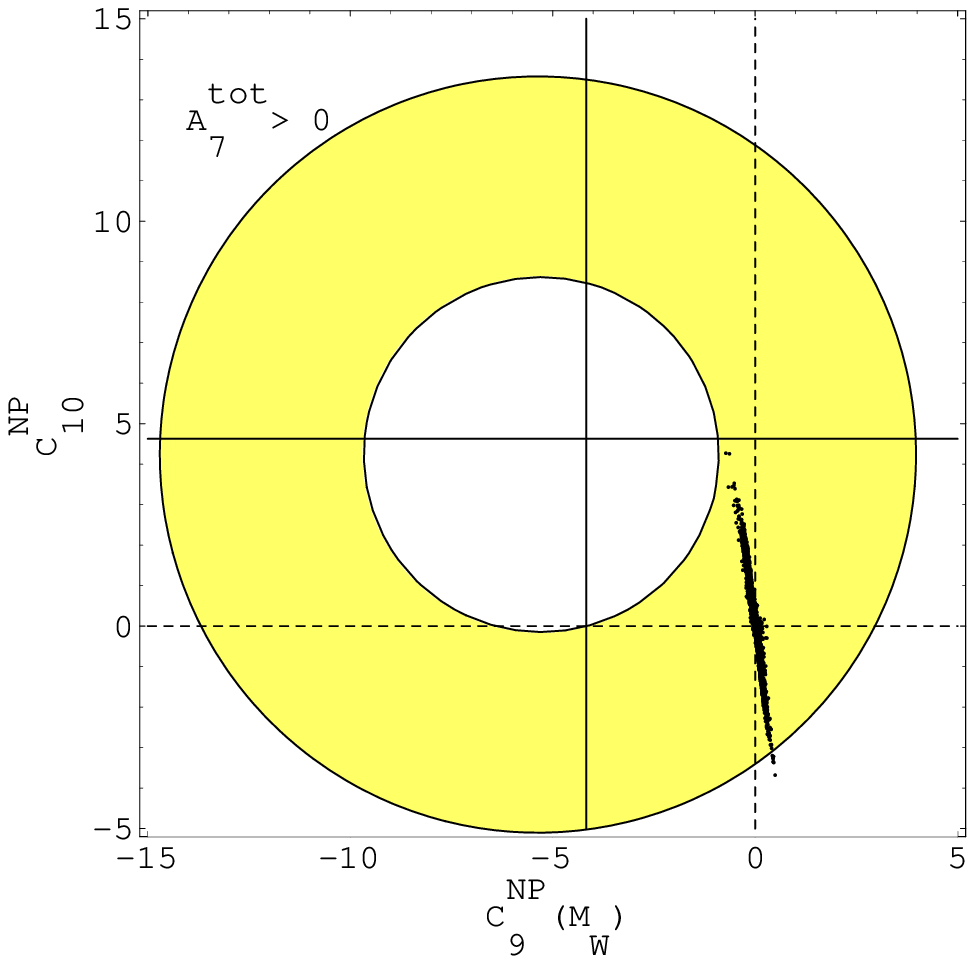,width=0.4\linewidth}
\caption{\it {\bf NNLO Case.} Superposition of all the
constraints. The plots correspond to the $A_7^{\rm tot}(2.5 \;
\gev)<0$ and $A_7^{\rm tot}(2.5 \; \gev) >0$ case, respectively. The
points are obtained by means of a scanning over the EMFV parameter
space and requiring the experimental bound from $B\to X_s \g$ to be
satisfied.}
\label{fig:total}
\end{center}
\end{figure}

In this section we compute, in the $[C_9^{NP} (\mu_W),C_{10}^{NP}]$
plane, the bounds implied by the experimental results given in
Eqs.~(\ref{bkmmexp})--(\ref{bseeexp}). The results are summarized in
Figs.~\ref{fig:seeNLO}--\ref{fig:total}. In each figure we focus on a
different experimental bound and the two plots shown in these figures
correspond respectively to the $A_7^{\rm tot}$-negative and $A_7^{\rm
tot}$-positive solutions just discussed. Within each plot we then vary
$A_7^{\rm tot}$ in the allowed range [given in
Eqs.~(\ref{a7lim})]. The present bounds impact more strongly the
decays $B\to (X_s,K^*) e^+ e^-$, for which the branching ratios are
larger due to the smallness of the electron mass. On the other hand,
the decays $B\to K \ell^+ \ell^-$ do not show any enhancement in the
low-$\hat{s}$ region and hence they are practically the same for the
dielectron and dimuon final states. Hence, the bounds for the
$Ke^+e^-$ and $K\m^+\m^-$ cases are presented in the same plot. In
\fig{fig:total} we combine all the bounds in a single plot. Note that
the overall allowed region is driven by the constraints emanating from
the decays $B\to X_s e^+ e^-$ and $B\to K \mu^+ \mu^-$.

\begin{figure}[H]
\begin{center}
\epsfig{file=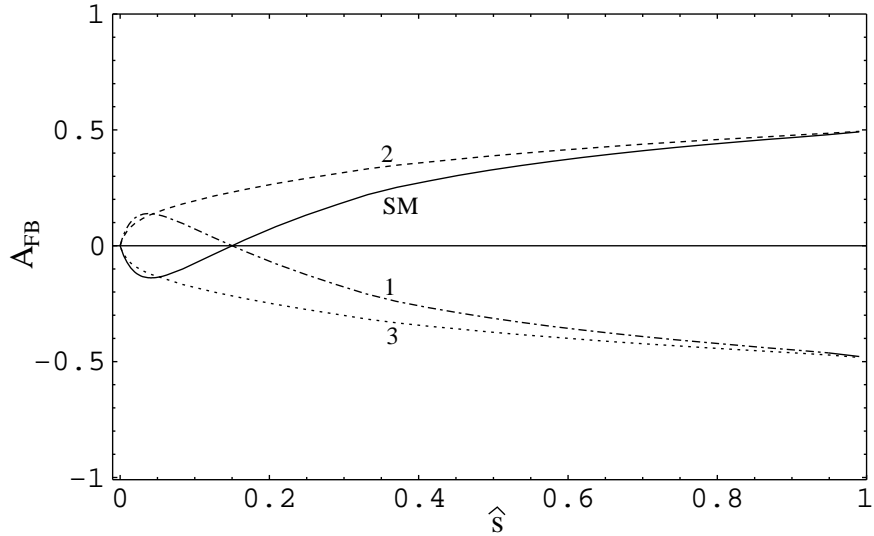,width=0.7\linewidth}
\caption{\it Differential Forward--Backward asymmetry for the decay
$B\to X_s \ell^+ \ell^-$. The four curves correspond to the points 
indicated in \fig{fig:total}.}
\label{fig:afb}
\end{center}
\end{figure}

Some comments on the results shown in these figures are in order:
\bit
\item From the comparison of
Figs.~\ref{fig:seeNLO}~and~\ref{fig:seeNNLO}, the importance of
performing the analysis using the NNLO precision clearly emerges. In
\fig{fig:seeNLO} we used the NLO precision ( see for instance in
\rf{CMW}). In this approximation we have to drop all the finite
corrections of order $\a_s$ (that is all the terms with the
superscript $(1)$) and the functions $F_i^{(j)}$, $\o_7$ and
$\o_{79}$; we retain the $\o_9$ term in $\tilde C_9^{\rm eff}$ but
drop the corresponding one in $\tilde C_{10}^{\rm eff}$. The impact of
switching on all these corrections is to lower sizably the branching
ratios (this happens both in the full and partial NNLO scenarios
previously discussed).  As a result, the strong constraints on the new
physics Wilson coefficients resulting from the NLO analysis are
softened by the inclusion of the NNLO corrections.
\item In \fig{fig:total} we identify four regions still allowed by the
constraints on the branching ratios that present very different
forward--backward asymmetries. In \fig{fig:afb} we show the shape of
the FB asymmetry spectrum for the SM and other three sample
points. The distinctive features are the presence or not of a zero and
global sign of the asymmetry. A rough indication of the FB asymmetry
behavior is thus enough to rule out a large part of the parameter
space that the current branching ratios can not explore.
\item For the decays $B\to K \mu^+ \mu^-$ and $B \to K e^+ e^-$, a
measurement is now at hand which we have already listed. The BELLE
collaboration has combined these branching ratios, getting ${\cal B}(
B\to K \ell^+ \ell^-)=0.75^{+0.25}_{-0.21} \pm 0.09 \times 10^{-6}$
\cite{bellebsll}. In showing the constraints in \fig{fig:kllNNLO} from
$B \to K \ell^+ \ell^-$, we have used this measurement to get the
following bounds:
\beq 
0.38 \; \times 10^{-6} \leq \branch (B\to K
\ell^+ \ell^-) \leq 1.2 \; \times 10^{-6} \; {\rm at} \; \cl{90} \; .
\label{BKLLBR}
\eeq 
\eit 
Concerning the upper bound, $1.2 \times 10^{-6}$, we note that
currently a discrepancy exists between the BELLE \cite{bellebsll} and
the BABAR \cite{babarbsll} results, with the latter reporting an upper
limit ${\cal B}(B \to K \ell^+ \ell^-) < 0.5 \times 10^{-6}$ (at 90\%
C.L.) conflicting mildly with the BELLE measurements. However, this
could just represent the vagaries of statistical fluctuations, and
hopefully this apparent mismatch will be soon resolved with more data.
Note that the branching ratio for $B \to K \ell^+ \ell^-$ is bounded both
from above and below, resulting in carving out an inner region in the
$(C_9^{\rm NP}(\m_W),C_{10}^{\rm NP})$ plane. 

At the end of this section we present the numerical expressions for
the inclusive branching ratios integrated over the low-$\hat s$ region
only where the full NNLO calculation is at hand. According to the
Belle analysis presented in \rf{bellebsll} we choose the integration 
limits as follows:
\bea
B\to X_s e^+ e^-: \; \left( 0.2\; \gev \over m_b \right)^2 \leq \hat s \leq 
      \left( M_{J/\Psi} - 0.6 \; \gev \over m_b \right)^2 \, , \\
B\to X_s \m^+ \m^-: \; \left( 2 m_\mu \over m_b \right)^2 \leq \hat s \leq 
      \left( M_{J/\Psi} - 0.35 \; \gev \over m_b \right)^2 \; .
\eea
The integrated branching ratios have the following form:
\bea
\branch (B\to X_s \ell^+ \ell^-) &=& 10^{-6} \times \Big[
       a_1 + a_2 \; |A_7^{\rm tot}|^2 + a_3 \; (|C_9^{\rm NP}|^2 +
 |C_{10}^{\rm NP}|^2) \nn \\
&&  + a_4 \; \Re A_7^{\rm tot} \; \Re C_9^{\rm NP} + a_5 \; \Im A_7^{\rm tot}
 \; \Im C_9^{\rm NP} 
    + a_6 \;  \Re A_7^{\rm tot} \nn \\
&&  +  a_7 \; \Im  A_7^{\rm tot} + a_8 \; \Re C_9^{\rm NP} +  a_9 \;
 \Im C_9^{\rm NP} + a_{10} \; \Re C_{10}^{\rm NP} \Big]\, , 
\eea
where the numerical value of the coefficients $a_i$ are given in
Table~\ref{table:numcoeff} for $\ell=e,\; \mu$. For the integrated
branching ratios in the SM we find:
\bea
\branch (B\to X_s e^+ e^-) &=& \left(2.47 \pm 0.40 \right) \times 10^{-6} \;\;\; 
        (\d \branch_{X_see} = \pm 16 \%)\; , \\
\branch (B\to X_s \m^+ \m^-) &=& \left(2.75 \pm 0.45 \right) \times 10^{-6} \;\;\; 
        (\d \branch_{X_s\m\m} = \pm 16 \%) \; . 
\eea
\begin{table}[htb]
\begin{center}
\begin{tabular}{|c|c|c|c|c|c|c|c|c|c|c|}\hline\hline
$\ell$&$a_1$&$a_2$&$a_3$&$a_4$&$a_5$&$a_6$&
$a_7$&$a_8$&$a_9$&$a_{10}$ \\ \hline
$e$ & 1.9927 & 6.9357 & 0.0640 & 0.5285 & 0.6574 & 0.2673 &
 -0.0586 & 0.4884 & 0.0095 & -0.5288 \\ \hline
$\m$& 2.3779 & 6.9295 & 0.0753 & 0.6005 & 0.7461 & 0.5955 &
-0.0600 & 0.5828 & 0.0102 & -0.6225 \\ \hline
\end{tabular}
\end{center}
\caption{Numerical values of the coefficients $a_i$ (evaluated at
$\mu_b=5\; \gev$) for the decays $B\to X_s \ell^+ \ell^-$ ($\ell=e,\;
\m$). $A_7^{\rm tot}$ is computed at $\mu_b=5\; \gev$ while $C_9^{\rm
NP}$ at $\mu_W = m_W$ ($C_{10}^{\rm NP}$ is scale independent). We use
the full NNLO calculation which is available only in the low--$\hat s$
region. The actual ranges for the integrations are chosen according
to the Belle analysis presented in \rf{bellebsll}. They are $s \in [4
m_\m^2 , (M_{J/\Psi} -0.35\; \gev)^2]$ for the $X_s \m^+\m^-$ and $s
\in [(0.2\; \gev)^2, (M_{J/\Psi} -0.60\; \gev)^2]$ for the $X_s e^+
e^-$ modes.}
\label{table:numcoeff}
\end{table}
\section{Analysis in supersymmetry}
\label{MIA} 
In this section we analyze the impact of the $b\to s\g$ and $b\to s
\ell^+ \ell^-$ experimental constraints on several supersymmetric
models. We will first discuss the more restricted framework of the
minimal flavour violating MSSM, and then extend the analysis to more
general models in which new SUSY flavour changing couplings are
allowed to be non--zero for which we will adopt the so--called mass
insertion approximation (MIA)~\cite{mia1,mia2}.
\subsection{Minimal Flavour Violation}
\label{MFV}
As already known from the existing literature (see for instance
\rf{LMSS}), minimal flavour violating (MFV) contributions are
generally too small to produce sizable effects on the Wilson
coefficients $C_9$ and $C_{10}$. In the MFV scheme all the genuine new
sources of flavour changing transitions other than the CKM matrix are
switched off, and the low energy theory depends only on the following
parameters: $\m$, $M_2$, $\tan \b$, $M_{H^\pm}$, $M_{\tilde t_2}$ and
$\theta_{\tilde t}$ (see Appendix~\ref{stopchargino} for a precise
definition of the various quantities).  Scanning over this parameter
space and taking into account the lower bounds on the sparticle masses
($M_{\tilde t_2} \geq 90 \; \gev$, $M_{\chi_i^\pm}\geq 90 \; \gev $) as
well as the $b\to s \g$ constraint given in \eq{bsgexp}, we derive the
ranges for the new physics contributions to $C_9$ and $C_{10}$. In
order to produce bounds that can be compared with the model
independent allowed regions plotted in \fig{fig:total}, we divided the
surviving SUSY points in two sets, according to the sign of $A_7^{\rm
tot}$. Scanning over the following parameter space
\beq
\cases{ 
M_{\tilde t}   =   90\; \gev \div 1 \; \tev \cr
\theta_{\tilde t}   =   -\p/2 \div \p/2 \cr
\tan \b = 2.3 \div 50 \cr
\mu   =  -1 \; \tev \div 1 \; \tev \cr
M_2  =  0 \div 1 \; \tev \cr
M_{H^{\pm}}  =  78.6 \; \gev \div 1 \; \tev \cr
M_{\tilde \n}   \geq   50 \; \gev    \cr}
\label{mfvps}
\eeq
we find that the allowed $C_9$ and $C_{10}$ ranges are:
\bea
A_7^{\rm tot}<0 \Rightarrow \cases{C_9^{MFV}(\m_W) \in [-0.2, 0.4]\, , \cr  
                                   C_{10}^{MFV}    \in [-1.0, 0.7]}\, . \\
A_7^{\rm tot}>0 \Rightarrow \cases{C_9^{MFV}(\m_W) \in [-0.2, 0.3]\, , \cr  
                                   C_{10}^{MFV}    \in [-0.8, 0.5] \; .}
\eea
We stress that the above discussion applies to any supersymmetric model
with flavour universal soft-breaking terms, such as
minimal supergravity MSSM and gauge-mediated supersymmetry
breaking models. Beyond-the-SM flavour violations in such models are
induced only via renormalization group running, and are tiny.
Hence, they can be described by MFV models discussed in this paper.

\begin{figure}[H]
\begin{center}
\epsfig{file=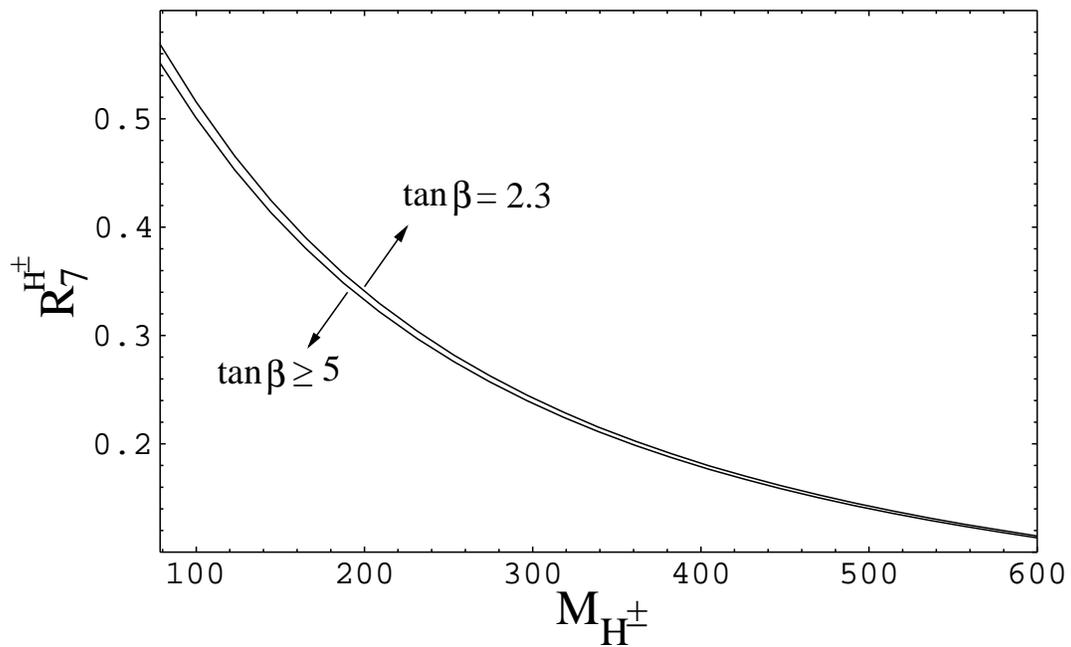,width=0.85\linewidth}
\caption{\it Dependence of $R_7^{H^\pm} (\m_b) \equiv C_7^{H^\pm}
(\m_b) / C_7^{\rm SM} (\m_b)$ on the mass of the charged Higgs.}
\label{fig:r7h}
\end{center}
\end{figure}

\begin{figure}[H]
\begin{center}
\epsfig{file=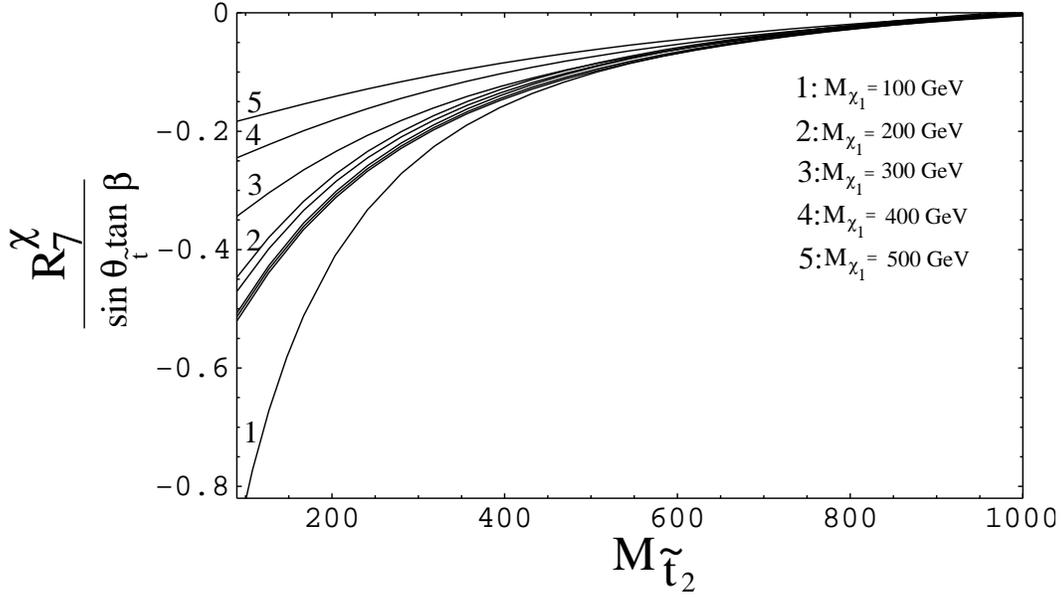,width=0.85\linewidth}
\caption{\it Dependence of $R_7^{\chi} (\m_b) \equiv C_7^{\chi} (\m_b)
/ C_7^{\rm SM} (\m_b)$ on the mass of the lightest stop in MFV models.
The chargino contribution is essentially proportional to $\sin
\theta_{\tilde t} \tan \b$ for not too small $\sin \theta_{\tilde t}$.
For the set of curves 2 we show the variation due to several choices of
$\theta_{\tilde t}$ and $\tan \beta$. The thick bunch of lines is
obtained for $(\sin \theta_{\tilde t},\tan \beta)=(0.025,40)$,
$(0.05,20)$, $(0.1,10)$, $(0.2,5)$, $(.5,2)$ for which 
$\sin \theta_{\tilde t} \tan \beta = 1$. The two thin curves 
correspond to $(\sin \theta_{\tilde t},\tan \beta)=(0.1,20)$ and  
$(.1,40)$, with the product $\sin \theta_{\tilde t} \tan
\beta$ having a value 2 and 4, respectively.}
\label{fig:r7ch}
\end{center}
\end{figure}

Before finishing this subsection and starting our discussion on models
with new flavour changing interactions, let us show in more detail the
impact of $b\to s \gamma $ on MFV models. The scatter plot presented
in Fig.~\ref{bsg} is obtained varying the MFV SUSY parameters
according to the above ranges and shows the strong correlation between
the values of the Wilson coefficients $C_7$ and $C_8$. In fact, the
SUSY contributions to the magnetic and chromo--magnetic coefficients
differ only because of colour factors and loop-functions. In
Figs.~\ref{fig:r7h} and \ref{fig:r7ch} we present the dependence of
the charged Higgs and chargino contributions to $C_7$ on the relevant
mass parameters (that are the charged Higgs mass for the former and
the lightest chargino and stop masses for the latter). Note that we
plot the SUSY Wilson coefficients at the scale $\m_b$ normalized to
the SM values. In the chargino case we are able to exploit the
$\theta_{\tilde t}$ and $\tan \b$ dependence since (for non negligible
values of the stop mixing angle) the chargino contribution is
essentially proportional to $\sin \theta_{\tilde t} \tan \b$. Indeed,
the various curves are very stable with respect to variations of $\tan
\beta$ and $\theta_{\tilde t}$ according to the ranges specified in
Eq.~(\ref{mfvps}); in order not to complicate unnecessarily the figure
we show the actual spread for the $M_{\chi} = 200 \; \gev$ case only
(note that according to the above discussion we require
$|\theta_{\tilde t}| \geq 0.01$). In order to show the full strength
of these figures let us entertain a scenario in which $C_7$ has the
same sign as in the SM. In this situation large contributions to $C_7$
are completely ruled out.  This means that, looking at
Figs.~\ref{fig:r7h} and \ref{fig:r7ch}, it is possible to obtain lower
bounds on some SUSY particles. Note that \fig{fig:r7ch} has very
strong consequences. Assuming for instance $M_{\tilde t_2} = M_{\chi}=
500\; \gev$ we see that the ratio $R_7^{\chi}/(\sin \theta_{\tilde t}
\tan \b)$ is of order 0.2. If we then allow for larger values of the
stop mixing angle and of $\tan \b$, the contribution can easily
violate the $b\to s\g $ constraint by more than one order of magnitude
(e.g. for $\sin \theta_{\tilde t} = 0.5 $ and $\tan \b=50$ we obtain
something of order 6 that is orders of magnitude above the current
limit).

\subsection{Gluino contributions}
\label{gluino}
Gluino contributions to $C_9$ and $C_{10}$ are governed by mass
insertions in the down squark mass matrix. From the analysis presented
in \rf{LMSS} we see that the dominant diagrams involve the parameter
$(\d^d_{23})_{LL}$ and that large deviations from the SM are unlikely.
The impact of $(\d^d_{23})_{LR}$ is negligible for the following two
reasons. First of all, contributions to either $C_9$ or $C_{10}$ are
obtained by $b_L \to s_L$ transitions and LR insertions can therefore
enter only at the second order in the mass insertion expansion. More
importantly, the insertion $(\d^d_{23})_{LR}$ gives a contribution to
the coefficient $C_7$ that is two orders of magnitude bigger than the
SM one. The bottom line of this discussion is that $(\d^d_{23})_{LR}$
contributions to the semileptonic Wilson coefficients are extremely
suppressed. Moreover, there are no gluino box diagrams and the
$\gamma$--penguins are enhanced with respect to the $Z$--ones so that
only contributions to $C_9$ are non vanishing. Their explicit
expression is (see \rf{LMSS} for the analytical equations):
\beq 
C_9^{\tilde g,MI}=-0.93 \left( 250 \gev \over M_{\tilde q}\right)^2 
{f_8^{MI} (x_{\tilde g\tilde q}) \over 1/3} (\d^d_{23})_{LL}\, ,
\eeq
where $x_{\tilde g\tilde q}=M_{\tilde g}^2/ M_{\tilde q}^2$, the
$f_8^{MI} (x)$ loop-function is always smaller than $1/3$ and can be
found in Appendix~\ref{loopfunctions}. The situation is thus similar
to the MFV case and the same conclusions hold.
\subsection{Chargino contributions: Extended--MFV models}
\label{EMFV}
A basically different scenario arises if chargino--mediated penguin
and box diagrams are considered. As can be inferred by Table~4 in
\rf{LMSS}, the presence of a light $\tilde t_2$ generally gives rise
to large contributions to $C_9$ and especially to $C_{10}$. In the
following, we will concentrate on the so--called Extended MFV (EMFV)
models that the two of us described in \rf{AL} and that we will
briefly summarize below. In these models we can fully exploit the
impact of chargino penguins with a light $\tilde t$ still working with
a limited number of free parameters.

EMFV models are based on the heavy squarks and gluino assumption.  In
this framework, the charged Higgs and the lightest chargino and stop
masses are required to be heavier than $100 \; \gev$ in order to
satisfy the lower bounds from direct searches.  The rest of the SUSY
spectrum is assumed to be almost degenerate and heavier than $1 \;
\tev$. The lightest stop is almost right--handed and the stop mixing
angle (which parameterizes the amount of the left-handed stop $\tilde
t_L$ present in the lighter mass eigenstate) turns out to be of order
$O(m_W / M_{\tilde q}) \simeq 10\%$; for definiteness we will take
$|\theta_{\tilde t}| \leq \p /10$. The assumption of a heavy ($\ge 1$
TeV) gluino totally suppresses any possible gluino--mediated SUSY
contribution to low energy observables. Note that even in the presence
of a light gluino (i.e. $M_{\tilde g} \simeq O(300 \; \gev)$) these
penguin diagrams remain suppressed due to the heavy down squarks
present in the loop. In the MIA approach, a diagram can contribute
sizably only if the inserted mass insertions involve the light
stop. All the other diagrams require necessarily a loop with at least
two heavy ($\geq 1 \; \tev$) squarks and are therefore automatically
suppressed. This leaves us with only two unsuppressed flavour changing
sources other than the CKM matrix, namely the mixings $\tilde u_L -
\tilde t_2$ (denoted by $\d_{\tilde u_L \tilde t_2}$) and $\tilde c_L
- \tilde t_2$ (denoted by $\d_{\tilde c_L \tilde t_2}$).  We note that
$\d_{\tilde u_L \tilde t_2}$ and $\d_{\tilde c_L \tilde t_2}$ are mass
insertions extracted from the up--squarks mass matrix after the
diagonalization of the stop system and are therefore linear
combinations of $(\d_{13})^U_{LR}$, $(\d_{13})^U_{LL}$ and of
$(\d_{23})^U_{LR}$, $(\d_{23})^U_{LL}$, respectively. The insertions
relevant to our discussion are normalized as follows:
\beq
\label{delta}
\d_{\tilde u(\tilde c)_L \tilde t_2} 
\equiv {M^2_{\tilde u(\tilde c)_L \tilde t_2}
\over M_{\tilde t_2} M_{\tilde q}} { |V_{td(s)}| \over V_{td(s)}^*} 
\; .  
\eeq 
The phenomenological impact of $\delta_{\tilde t_2 \tilde u_L}$ has
been studied in \rf{AL} and its impact on the $b\to s \g$ and $b\to s
\ell^+ \ell^-$ transitions is indeed negligible. Therefore, we are
left with the MIA parameter $\delta_{\tilde t_2 \tilde c_L}$ only.
Thus, the SUSY parameter space that we have to deal with is: $\mu$,
$M_2$, $\tan \beta$, $M_{\tilde t_2}$, $\sin \theta_{\tilde t}$,
$M_{H^\pm}$, $M_{\tilde \nu}$ and $\delta_{\tilde t_2 \tilde c_L}$.

The explicit expressions for the mass insertion contributions to
the Wilson coefficients $C_{7}$ -- $C_{10}$ are summarized in 
Appendix~\ref{MIAcoefficients}. 

In order to explore the region in the $[C_9^{NP},C_{10}^{NP}]$ plane
(where $C_{9,10}^{NP}$ are the sum of MFV and MI contributions and are
explicitly defined in Appendix~\ref{MIAcoefficients}) that is
accessible to these models, we performed a high statistic scanning
over the following EMFV parameter space requiring each point to
survive the constraints coming from the sparticle masses lower bounds
and $b\to s \g$:
\beq
\cases{ 
M_{\tilde t}   =   90\; \gev \div 1 \; \tev \cr
\theta_{\tilde t}   =   -\p/10 \div \p/10 \cr
\tan \b = 2.3 \div 50 \cr
\mu   =  -1 \; \tev \div 1 \; \tev \cr
M_2  =  0 \div 1 \; \tev \cr
M_{H^{\pm}}  =  78.6 \; \gev \div 1 \; \tev \cr
M_{\tilde \n}   \geq   50 \; \gev    \cr
\delta_{\tilde t_2 \tilde c_L} = -1 \div 1 \; . \cr}
\eeq
The surviving points are shown in \fig{fig:total} together with the
model independent constraints. Note that these SUSY models can account
only for a small part of the region allowed by the model independent
analysis of current data. We stress that in our numerical analysis
reported here, we have used the integrated branching ratios to put
constraints on the effective coefficients. This procedure allows
multiple solutions, which can be disentangled from each other only
with the help of both the dilepton mass spectrum and the
forward-backward asymmetry. Only such measurements would allow us to
determine the exact values and signs of the Wilson coefficients $C_7$,
$C_9$ and $C_{10}$. 

\section{Summary}
\label{Summary}
We have presented theoretical branching ratios for the rare $B$ decays
$B \to X_s \ell^+ \ell^-$ and $B \to (K,K^*) \ell^+ \ell^-$,
incorporating the NNLO contributions in the former and partial NNLO
improvements in the latter. This has allowed us to carry out a
theoretical analysis of the radiative decays $B \to X_s \g$ and the
mentioned semileptonic decays to the same order in $\alpha_s$. In
addition, we have included the leading power corrections in $1/m_b$
and $1/m_c$ in the inclusive decays. The dilepton invariant mass
spectrum is calculated in the NNLO precision in the low dilepton
invariant mass region, $\hat{s} < 0.25$.  The spectrum for $\hat{s} >
0.25$ calculated to the same theoretical accuracy is not yet
available. We estimate the spectrum in this range from the known
partial NNLO, by noting that the dilepton mass spectrum in the full
NNLO is close to the partial NNLO spectrum in the range $\hat{s} <
0.25$ for the choice of the scale $\m_b=2.5$ GeV. Following this
observation, we use the partial NNLO spectrum with this scale to
estimate the central value of the full NNLO spectrum for $\hat{s} >
0.25$.  The branching ratios in the NNLO accuracy in the SM are
calculated to have the values $\branch (B\to X_s e^+ e^-)=(6.89 \pm
1.01) \times 10^{-6}$ and $\branch (B\to X_s \mu^+ \mu^-)=(4.15 \pm
0.7) \times 10^{-6}$. They are lower by typically 12\% and 20\%,
respectively, compared to their NLO estimates for the central values
of the input parameters, and are approximately a factor 2 to 4 away
from their respective experimental upper limits. Hence, current B
factory experiments will soon probe these decays at the level of the
SM sensitivity. In view of the fact that the dilepton mass spectrum is
calculated to the NNLO accuracy only for $\hat{s} < 0.25$, and the
long-distance effects are not expected to be dominant, we stress the
need to measure the inclusive decays $B \to X_s \ell^+ \ell^-$ in this
dilepton mass range. In fact, as shown in this paper, such a
measurement is theoretically as robust as the inclusive radiative
decay $B \to X_s \g$.

In the second part of this paper, we have used our improved
theoretical calculations to extract from the current data, listed in
Eqs.~(\ref{bsgexp}) -- (\ref{bseeexp}), the allowed ranges of the
effective Wilson coefficients $C_7(\mu)$ -- $C_{10}(\mu)$. In doing
this, we have first determined the ranges on the Wilson coefficients
$C_7(\m)$ and $C_{8}(\m)$ from $B \to X_s \g$ decay, and then
determined the allowed ranges of the coefficients $C_9^{NP}(\m_W)$ and
$C_{10}^{NP}$ (at 90\% C.L.). Since the decays $B \to K \ell^+ \ell^-$
are now measured by the BELLE collaboration, they carve out an inner
region in the ($C_9^{NP}(\m_W)$,$C_{10}^{NP}$) plane, allowed
previously.  Under the assumption that the SM-operator basis of the
effective Hamiltonian is sufficient to incorporate also the
beyond-the-SM physics effects, the analysis presented in this paper is
model independent. We find that all current data are consistent with
the SM. However, present experimental measurements allow considerable
room for beyond-the-SM effects, which we have worked out in specific
supersymmetric contexts. For this purpose, we have used the MFV model,
and an Extended-MFV model introduced in Ref.~\cite{AL}. The resulting
constraints on the supersymmetric parameters are worked out, in
particular on the charged Higgs mass, $M_{H^\pm}$, the lighter of the
two stop masses, $M_{\tilde{ t}_2}$, the ratio of the two Higgs vacuum
expectation values, $\tan \beta$, and the MIA parameter
$(\delta_{23})$. With more data, expected from the leptonic and
hadronic $B$ factories, these constraints will become either much more
stringent, pushing the supersymmetric frontier further, or else, more
optimistically, new data may lead to impeccable evidence for new
physics effects. We have illustrated this using the forward-backward
asymmetry in $B \to X_s \ell^+ \ell^-$ decays.

\section{Acknowledgments}
We would like to thank Wulfrin Bartel, Ed Thorndike, Mikihiko Nakao,
Hiro Tajima, Vera L\"uth and Howie Haber for helpful discussions and
communications on the data. We also thank Patricia Ball for a helpful
correspondence on the exclusive decay form factors. C.G. and
G.H. would like to thank the DESY theory group for warm hospitality
during their stay in Hamburg, where numerous stimulating discussion
regarding this work took place. E.L. acknowledges financial support
from the Alexander von Humboldt Foundation. The work of C. G. was
partially supported by Schweizerischer Nationalfonds. The work of
G. H. was supported by the Department of Energy, Contract
DE-AC03-76SF00515.

\appendix
\section{Stop and chargino mass matrices}
\label{stopchargino}
The $2\times 2$ stop mass matrix is given by
\beq
M^2_{\tilde t} = \pmatrix{ M^2_{\tilde t_{LL}} & M^{2}_{\tilde{t}_{LR}} \cr
                        M^{2*}_{\tilde{t}_{LR}} & M^{2}_{\tilde{t}_{RR}} \cr}
\, ,
\eeq
where
\bea
 M^{2}_{\tilde{t}_{LL}} &=& M^{2}_{\tilde q}
      + ( {1\over 2} - {2\over 3} \sin^{2}\theta_{W} ) \cos 2\beta \, m_Z^2 + m_{t}^{2} \enspace , \\
 M^{2}_{\tilde{t}_{RR}} &=& M^{2}_{\tilde q}
      + {2\over 3} \sin^{2}\theta_{W} \cos 2\beta \, m_{Z}^{2} + m_{t}^{2} \enspace , \\
 M^{2}_{\tilde{t}_{LR}} &=& m_{t} ( A_{t} - \mu^{*} \cot \beta )
\enspace .
\eea
The eigenvalues are given by
\beq
2 M^{2}_{\tilde t_1, \tilde t_2}
= ( M^{2}_{\tilde{t}_{LL}} + M^{2}_{\tilde{t}_{RR}} )
\pm \sqrt{ ( M^{2}_{\tilde{t}_{LL}} - M^{2}_{\tilde{t}_{RR}} )^{2}
         + 4 ( M^{2}_{\tilde{t}_{LR}} )^{2}}
\enspace ,
\eeq
with $M^2_{\tilde t_2} \le M^{2}_{\tilde t_1}$.
We parametrize the mixing matrix ${\mathcal R}^{\tilde{t}}$ so that
\beq
\pmatrix{ \tilde{t}_{1} \cr \tilde{t}_{2}} =
{\mathcal R}^{\tilde{t}}
\left(\begin{array}{c}
  \tilde{t}_{L} \\ \tilde{t}_{R}
\end{array}\right)
=
\left(\begin{array}{cc} 
 \cos \theta_{\tilde{t}} & +\sin \theta_{\tilde{t}} \\
  - \sin \theta_{\tilde{t}} &  \cos \theta_{\tilde{t}}
\end{array}\right)
\left(\begin{array}{c}
  \tilde{t}_{L} \\ \tilde{t}_{R}
\end{array}\right)
\enspace .
\eeq
The chargino mass matrix
\begin{equation}\label{charmass}
M^{\tilde{\chi}^{+}}_{\alpha\beta} =
\left(
\begin{array}{cc}
  M_2                        & m_{W} \sqrt{2} \sin\beta  \\
  m_{W} \sqrt{2} \cos\beta & \mu
\end{array}
\right) \, ,
\end{equation}
can be diagonalized by the bi-unitary transformation
\begin{equation}
\tilde U^{*}_{j\alpha} M^{\tilde{\chi}^{+}}_{\alpha\beta} \tilde V^{*}_{k\beta}
= M_{\tilde{\chi}_{j}^{+}} \delta_{jk}
\enspace ,
\end{equation}
where $\tilde U$ and $\tilde V$ are unitary matrices such that
$M_{\tilde{\chi}_{j}^{+}}$ are positive and
$M_{\tilde{\chi}_{1}^{+}} < M_{\tilde{\chi}_{2}^{+}}$.

\section{Wilson coefficients $C_{7}$ -- $C_{10}$ in EMFV models}
\label{MIAcoefficients}
In this appendix we collect the explicit expressions for the Wilson
coefficients $C_{7}$ -- $C_{10}$ in the mass insertion approximation.  The
conventions for the definition of the chargino mass matrix is
summarized in Appendix~\ref{stopchargino}, the normalization of the
mass insertion is given in \eq{delta} in the text, and the loop functions
encountered below can be found in Appendix~\ref{loopfunctions}.

\noindent
\bit 
\item Contributions to the magnetic and chromo--magnetic dipole
moment coefficients:
\eit
\bea
 C_{7,8}^{MI} &=& {\delta_{\tilde t_2 \tilde c_L} \over 6} \left| V_{cs}
               \over V_{ts}\right| 
            {m_W^2 \over M_{\tilde q}^2} {M_{\tilde t_2} \over M_{\tilde q}} \sum_{i=1}^2 V_{i1}^{}
             \Bigg[ 
     \left( \sin \theta_{\tilde t}V_{i1}^* -{m_t \cos \theta_{\tilde t} \over \sqrt{2} \sin \b m_W}V_{i2}^*
        \right)
            f_{1,3}^{\rm MI} (x_i,x_{\tilde t_2}) + \nn \\
        & & \sin  \theta_{\tilde t} U_{i2}^* 
            {\sqrt{2} M_{\chi_i} \over m_W \cos \b} f_{2,4}^{\rm MI}
         (x_i,x_{\tilde t_2}) \Bigg] \; .
\eea

\noindent The contributions to the semileptonic coefficients can be
divided in three classes: \bit
\item Photon mediated penguin diagrams:
\bea
 C_{10}^{MI,\g} & = & 0 \; , \\
 C_9^{MI,\g} & = & {1\over 9}\delta_{\tilde t_2 \tilde c_L}  
            \left| V_{cs} \over V_{ts}\right| 
            {m_W^2 \over M_{\tilde q}^2} {M_{\tilde t_2} \over M_{\tilde q}} \sum_{i=1}^2V_{i1}^{}  
     \left( {m_t \cos \theta_{\tilde t} \over \sqrt{2} \sin \b m_W}V_{i2}^* -
     \sin \theta_{\tilde t}V_{i1}^* \right)
      f_7^{\rm MI} (x_i,x_{\tilde t_2}) \; .
\eea 
\item Z mediated penguin diagrams:
\bea
 C_{10}^{MI,Z} &=& {\delta_{\tilde t_2 \tilde c_L} \over 4 \sin^2 \theta_W} 
                   \left| V_{cs} \over V_{ts}\right| {M_{\tilde t_2} \over M_{\tilde q}} 
                    \sum_{i,j,=1}^2 V_{i1}^{} \Bigg\{
                   \left( \sin \theta_{\tilde t}V_{j1}^*   - 
                    {m_t \cos \theta_{\tilde t} \over \sqrt{2}
                    \sin \b m_W} V_{j2}^* \right) \nn \\
   & & \hskip -2.2cm 
       \times \left( U_{i1}^* U_{j1}^{} {M_{\chi_i} M_{\chi_j} \over
        M_{\tilde q} M_{\tilde t_2}} j
        (x_i,x_j,x_{\tilde t_2}) +
        V_{i1}^* V_{j1}^{} {k (x_i,x_j,x_{\tilde t_2})\over 2 x_{\tilde t_2}} -
        \d_{ij} V_{i1}^{} V_{i2}^*  {k (x_i,x_{\tilde t_2},1)\over
         2 x_{\tilde t_2}} \right) \nn \\
  & & \hskip -2.2cm 
       - \sin \theta_{\tilde t} V_{j1}^* \; \d_{ij} V_{i1}^{} V_{i2}^*  {k (x_i,x_{\tilde t_2},1)\over
         2 x_{\tilde t_2}} \Bigg\} \; , \\
 C_9^{MI,Z} & = & (4 \sin^2 \theta_W-1) C_{10}^{MI,Z} \; .
\eea
\item Box diagrams with an internal sneutrino line:
\bea
 C_{10}^{MI,box} &=& {\delta_{\tilde t_2 \tilde c_L} \over \sin^2 \theta_W}
                 \left| V_{cs} \over V_{ts}\right| 
                 {m_W^2 \over M_{\tilde q}^2} {M_{\tilde t_2} \over M_{\tilde q}}\sum_{i,j=1}^2
                  |V_{i1}^{}|^2 V_{j1}^{} 
    \left( {m_t \cos \theta_{\tilde t} \over \sqrt{2} \sin \b m_W}V_{j2}^*
     - \sin \theta_{\tilde t}V_{j1}^* \right) \nn \\ 
     & &  \times d_2^{\rm MI} (x_i,x_j,x_{\tilde t_2},x_{\tilde 
     \nu}) \; ,\\C_9^{MI,box} &=& - C_{10}^{MI,box} \; .
\eea
\eit
The branching ratios for the various decays are obtained from
Eqs.~(\ref{effcoeff9})--(\ref{effcoeff10}) by means of the following
replacement:
\bea
C_{7,8}^{NP} & \rightarrow & C_{7,8}^{MFV} +  C_{7,8}^{MI} \; , \\
C_{9,10}^{NP} & \rightarrow & C_{9,10}^{MFV} +  C_{9,10}^{MI,\g} +
C_{9,10}^{MI,Z}+ C_{9,10}^{MI,box}\; ,
\eea
where the expressions for $C_i^{MFV}$ can be found in \rf{CMW}.

\section{Loop functions}
\label{loopfunctions}
The various loop functions introduced in Appendix \ref{MIAcoefficients} 
are listed below.
\begin{eqnarray}
  f_1 (x)  &=& {-7 + 12 x + 3 x^2 - 8 x^3 + 6 x (-2 + 3 x) \log x \over
  6 (x-1)^4}\; , \\
  f_2 (x) &=& {5 - 12x + 7x^2 - 2x(-2 + 3x)\log x \over 2 (x-1)^3} \; , \\
  f_3 (x) &=& {2 + 3 x - 6 x^2 + x^3 + 6 x \log x   \over 6 (x-1)^4}
\; ,  \\
  f_4 (x) &=& {-1 +x^2 - 2 x \log x   \over 2 (x-1)^3} \; , \\ 
  f_i^{MI} (x,y)     &=& { f_i (1/x) - f_i (y/x) \over x (1-y) } \;\;\;\;
 (i=1,2,3,4)  \; .
\end{eqnarray}

\begin{eqnarray}
  f_7 (x) &=& {52 - 153 x + 144 x^2 - 43 x^3 + 6 (6 - 9 x + 2 x^3) \log x
  \over 6 (x-1)^4} \; ,\\
  f_8 (x) &=& {2 - 9x + 18x^2 - 11x^3 + 6x^3\log x \over (x-1)^4} \; ,\\
  f_i^{MI} (x,y)     &=& { f_i (x) - f_i (x/y) \over 1-y}
\;\;\;\; (i=7,8)
   \; .
\end{eqnarray}

\begin{eqnarray}
j(x) &=& {x \log x \over x-1} \; , \; \;\; \;
j(x,y) = {j(x)-j(y) \over x-y} \; , \;\;\; \;
j(x,y,z) = {j(x,z)-j(y,z) \over x-y} \; , \\
k(x) &=& {x^2 \log x \over x-1} \; , \;\;\; \;
k(x,y) = {k(x)-k(y) \over x-y}\; , \;\;\; \;
k(x,y,z) = {k(x,z)-k(y,z) \over x-y} \; .
\end{eqnarray}

\begin{eqnarray}
d_2 (x,y,z,t) &=& -{1\over 4}\left[ {x \log x \over (x-y)(x-z)(x-t)} +
       (x\leftrightarrow y)+(x\leftrightarrow z)+(x\leftrightarrow
t)\right]\;
  , \\
d_2^{MI} (x,y,z,t) &=& { d_2 (x,y,1,t) - d_2 (x,y,z,t) \over 1 - z } \; .
\end{eqnarray}

\newpage


\begin{figure}[H]
\begin{center}
\epsfig{file=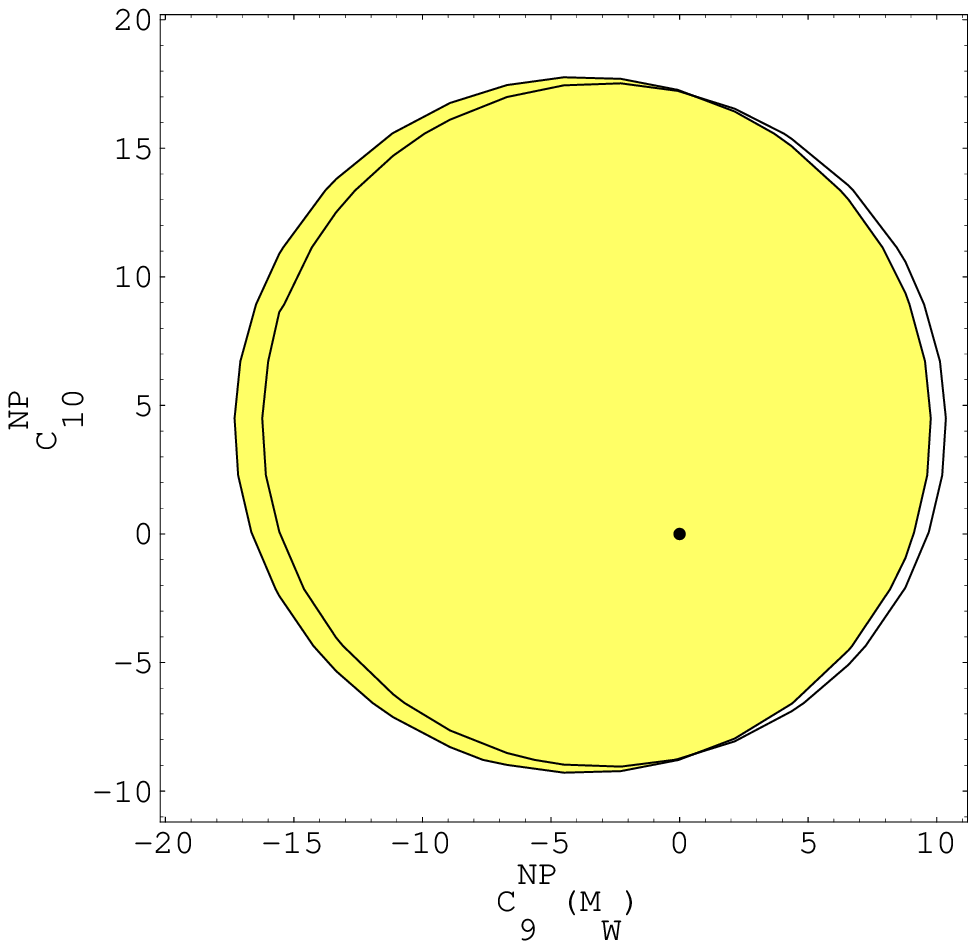,width=0.4\linewidth}
\hskip 0.5cm 
\epsfig{file=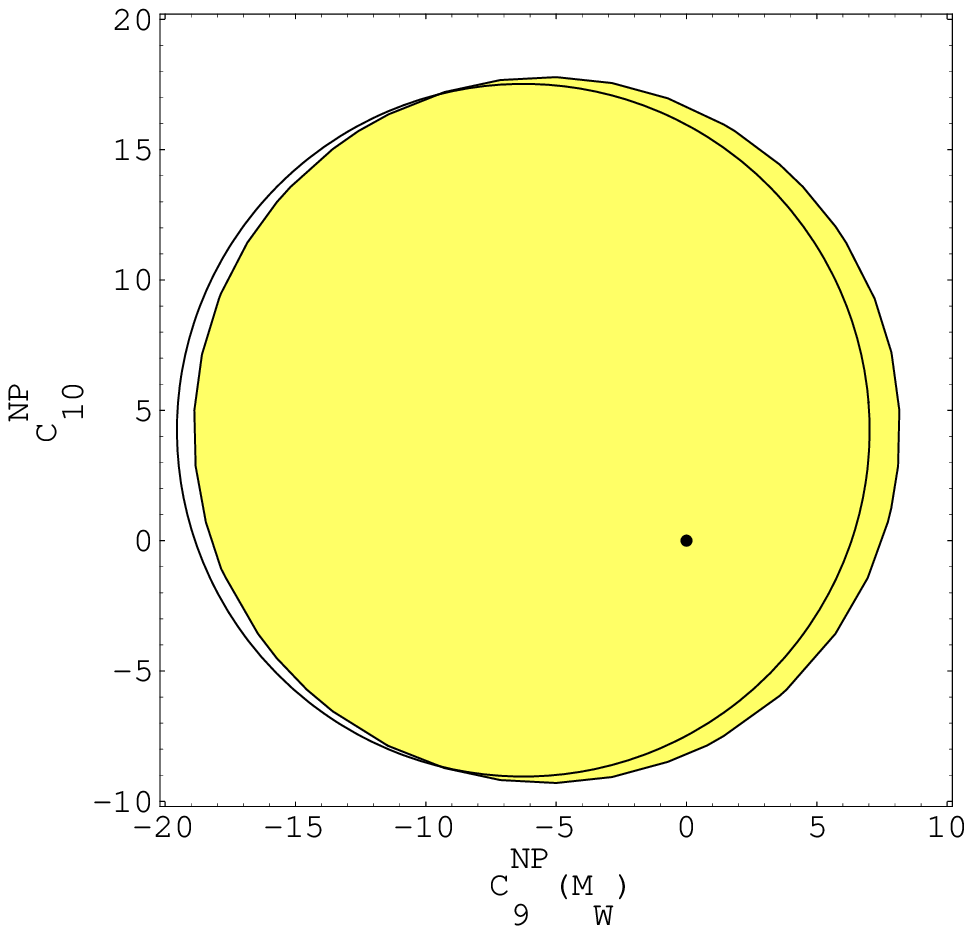,width=0.4\linewidth}
\caption{\it {\bf NNLO Case.} Constraints in the
$[C_9^{NP}(\mu_W),C_{10}^{NP}]$ plane that come from the BELLE
$\cl{90}$ upper limit $\branch (B\to X_s \mu^+ \mu^-) \leq 19.1 \times
10^{-6}$. See \fig{fig:seeNLO} for further details.}
\end{center}
\end{figure}

\begin{figure}[H]
\begin{center}
\epsfig{file=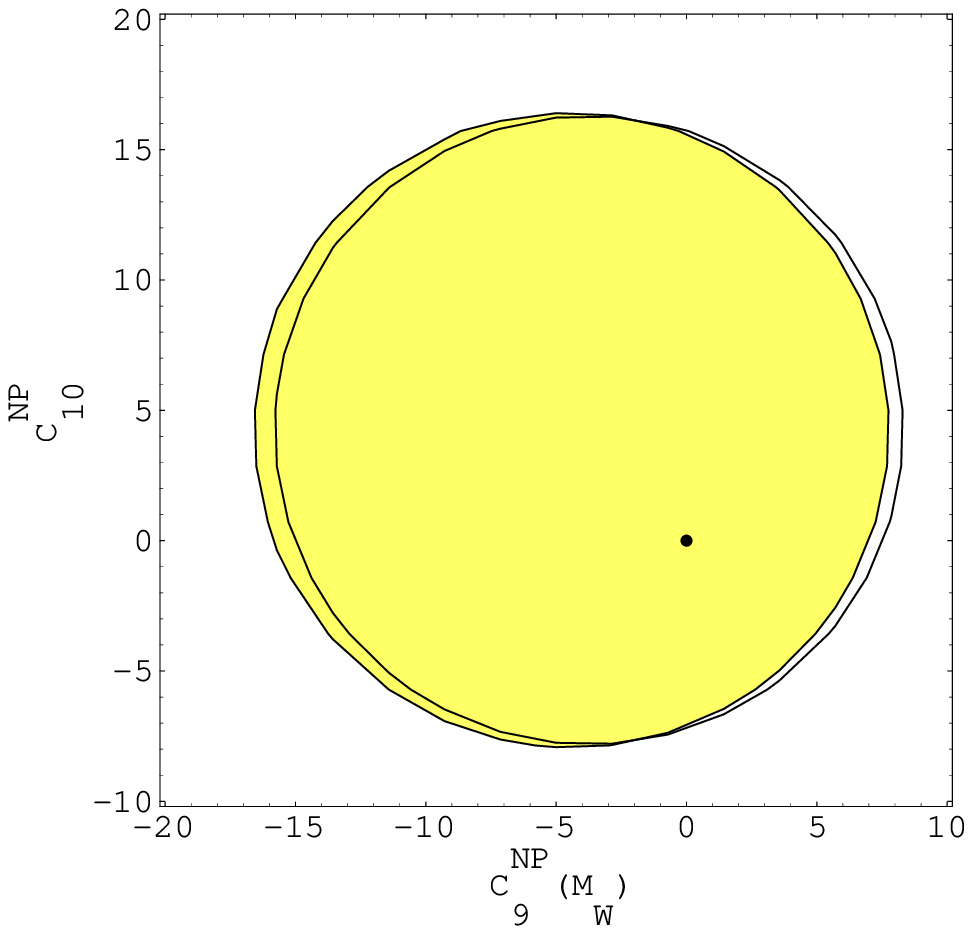,width=0.4\linewidth}
\hskip 0.5cm 
\epsfig{file=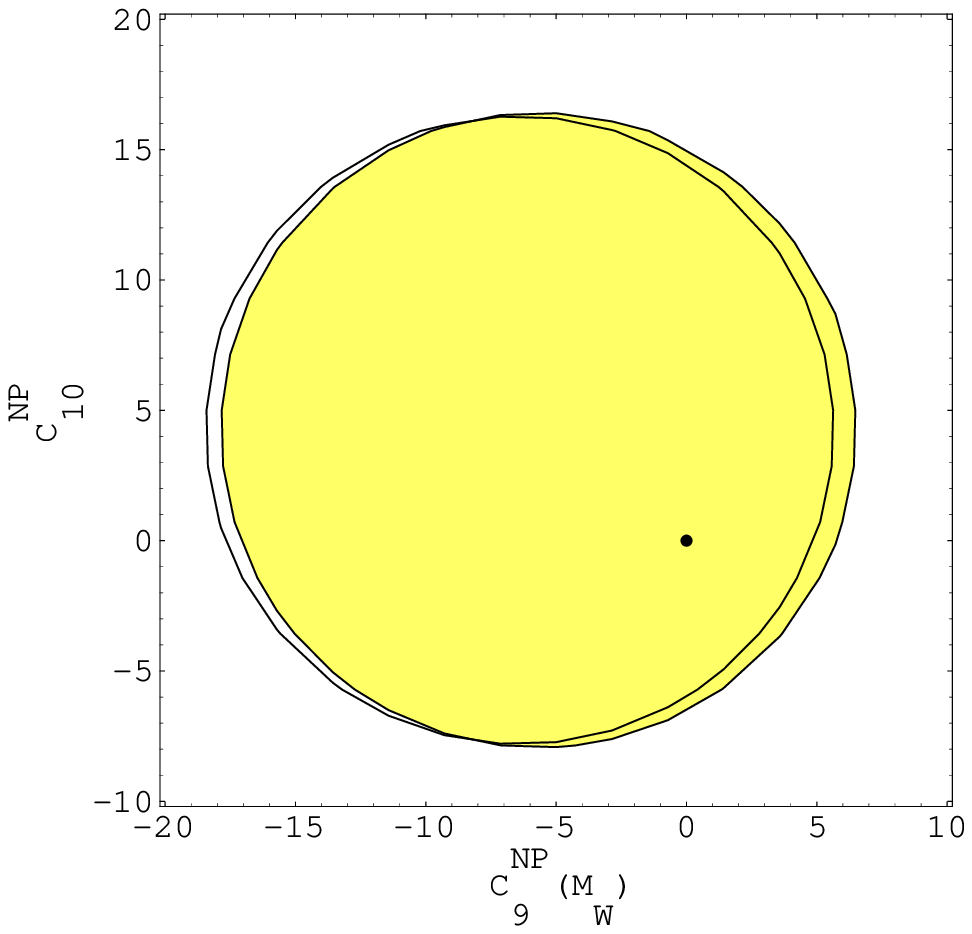,width=0.4\linewidth}
\caption{\it {\bf NNLO Case.} Constraints in the
$[C_9^{NP}(\mu_W),C_{10}^{NP}]$ plane that come from the $90\% CL$
BELLE constraint $\branch (B\to K^* \mu^+ \mu^-) \leq 3.0 \times 
10^{-6}$. See \fig{fig:seeNLO} for further details.}
\end{center}
\end{figure}

\begin{figure}[H]
\begin{center}
\epsfig{file=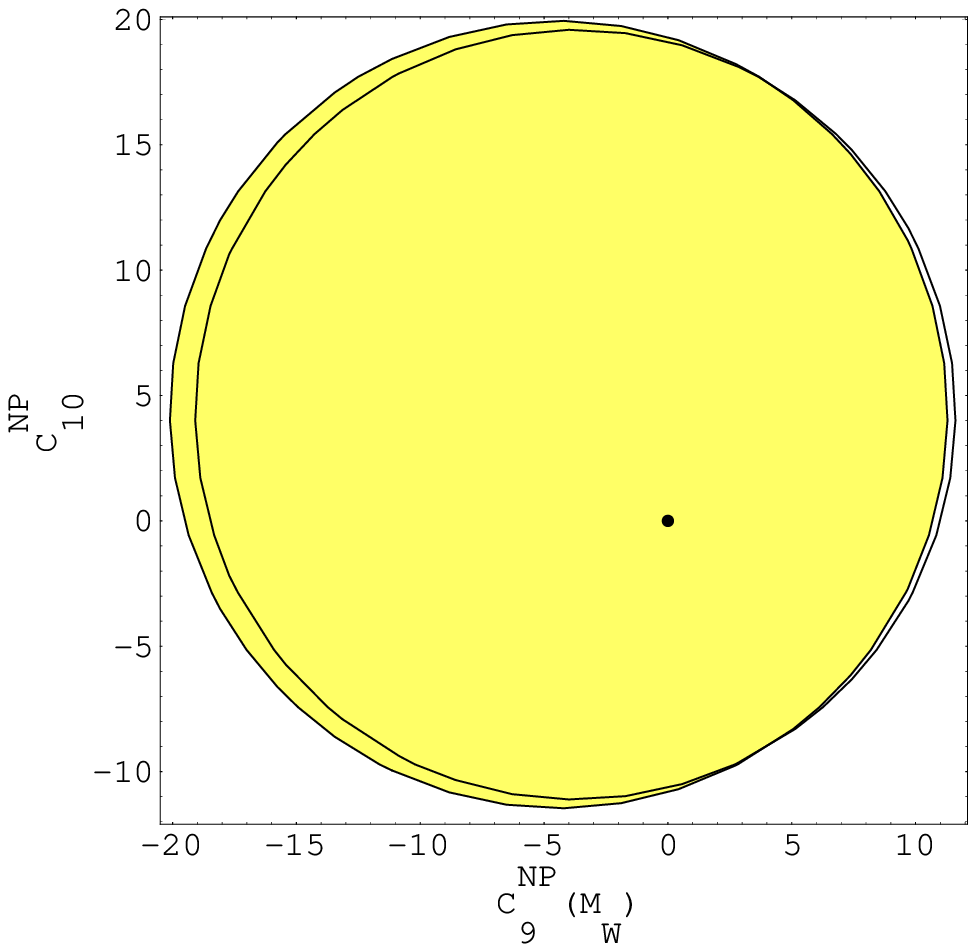,width=0.4\linewidth}
\hskip 0.5cm 
\epsfig{file=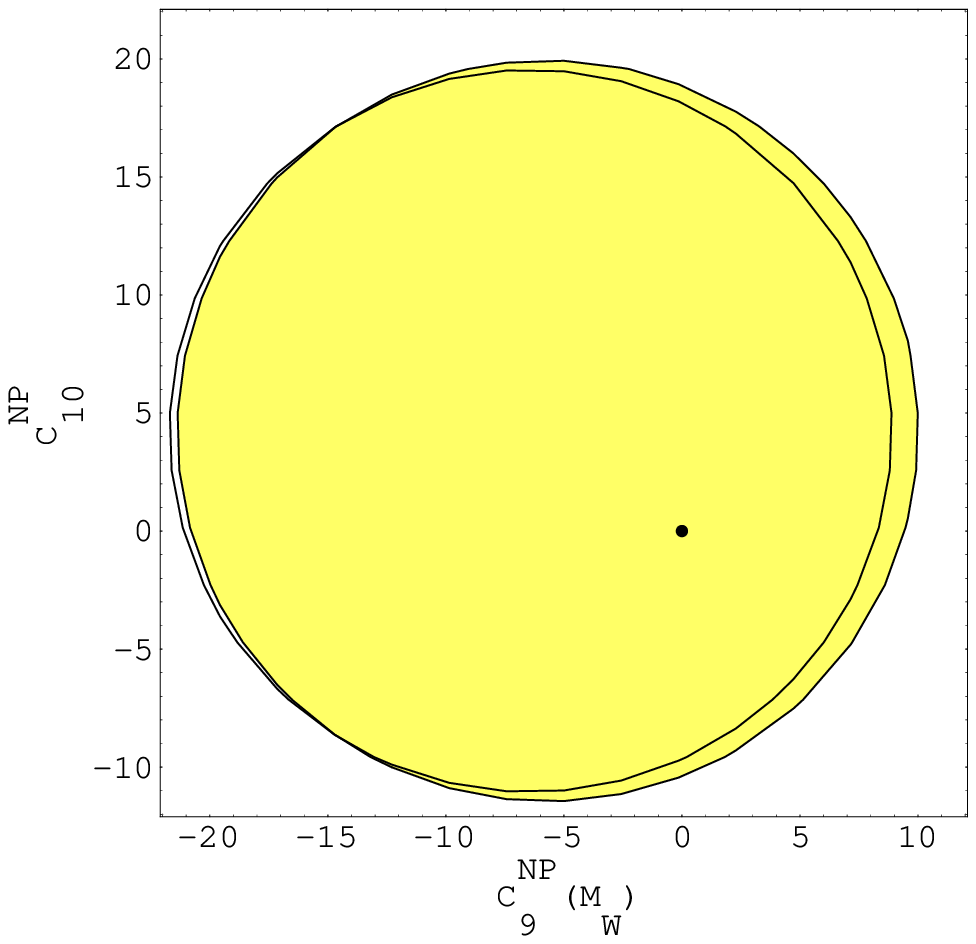,width=0.4\linewidth}
\caption{\it {\bf NNLO Case.} Constraints in the
$[C_9^{NP}(\mu_W),C_{10}^{NP}]$ plane that come from the $90\% CL$
BELLE constraint $\branch (B\to K^* e^+ e^-) \leq 5.1 \times 
10^{-6}$. See \fig{fig:seeNLO} for further details.}
\end{center}
\end{figure}

\end{document}